\RequirePackage{snapshot}
\documentclass[12pt]{iopart}

\usepackage[usenames,dvipsnames,svgnames,table]{xcolor}

\expandafter\let\csname equation*\endcsname\relax
\expandafter\let\csname endequation*\endcsname\relax
\usepackage{amsmath}
\usepackage{amssymb}
\usepackage{amsthm}
\usepackage{graphicx}
\usepackage{epstopdf}

\usepackage{amsmath}
\usepackage{amssymb}
\usepackage{amsthm}
\usepackage{graphicx}
\usepackage{caption}

\newcommand{\beq}{\begin{equation}}
\newcommand{\eeq}{\end{equation}}

\newcommand{\ba}{\begin{array}}
\newcommand{\ea}{\end{array}}

\newcommand{\bea}{\begin{eqnarray}}
\newcommand{\eea}{\end{eqnarray}}

\newcommand{\bc}{\begin{center}}
\newcommand{\ec}{\end{center}}

\newcommand{\bt}{\begin{tabular}}
\newcommand{\et}{\end{tabular}}

\def\bi{\begin{itemize}}
\newcommand{\ei}{\end{itemize}}

\newcommand{\bd}{\begin{description}}
\newcommand{\ed}{\end{description}}

\newcommand{\bp}{\begin{pmatrix}}
\newcommand{\ep}{\end{pmatrix}}

\newcommand{\p}{\partial}

\newcommand{\eg}{\mbox{e.g.}~}

\newcommand{\bk}{{\bf k}}

\newcommand{\bv}{{\bf v}}

\def\br{{\bf r}}

\newcommand{\bdd}{{\bf d}}

\newcommand{\ipro}[2]{\left<#1,#2\right>}

\newcommand{\eq}{\mbox{Eq.}}

\def\digamma{S}

\def\fC{C}

\begin{document}
\title{Adiabatic Dynamics of Edge Waves in Photonic Graphene}

\author{Mark J. Ablowitz$^1$, Christopher W. Curtis$^2$, and Yi-Ping Ma$^1$}
\address{$^1$Department of Applied Mathematics, University of Colorado, Boulder, Colorado 80309, USA\\
$^2$Department of Mathematics and Statistics, San Diego State University, San Diego, California 92182, USA}

\begin{abstract}
The propagation of localized edge modes in photonic honeycomb lattices, formed from an array of adiabatically varying periodic helical waveguides, is considered.  Asymptotic analysis leads to an explicit description of the underlying dynamics. Depending on parameters, edge states can exist over an entire period or only part of a period; in the latter case an edge mode can effectively disintegrate and scatter into the bulk.  In the presence of nonlinearity, a `time'-dependent one-dimensional nonlinear Schr\"odinger (NLS) equation describes the envelope dynamics of edge modes. When the average of the `time varying' coefficients yields a focusing NLS equation, soliton propagation is exhibited. For both linear and nonlinear systems, certain  long lived traveling modes with minimal backscattering are found; they exhibit properties of topologically protected  states.
\end{abstract}
\maketitle
\section{Introduction}
Recently there has been significant effort directed towards understanding the wave dynamics in photonic lattices arranged in a honeycomb structure cf. \cite{Segev1,Segev2,Segev3,Segev4,abl1,abl2,feffwein}. Due to the extra symmetry of the honeycomb lattice, Dirac points, or conical intersections between dispersion bands, exist. This is similar to what occurs in carbon-based graphene \cite{geim} where the existence of Dirac points
is a key reason for many of its exceptional  properties. Because of the correspondence with carbon-based graphene, the optical analogue is often termed `photonic graphene'. In lattices without edges the wave dynamics exhibit conical, elliptic, and straight line diffraction cf.  \cite{Segev1,Segev3,abl3,abl2}.  However when edges and a `pseudo-field' are present, remarkable
changes occur and long lived, persistent linear and nonlinear traveling edge waves with little backscatter appear.  These localized waves exhibit the hallmarks of topologically protected states, thus indicating photonic graphene is a topological insulator \cite{Kane2,Kane3,rechts2}.

Substantial attention has been paid to the understanding of edge modes in both condensed matter physics and optics.  Interest in such modes goes back to the first studies of the Quantum Hall Effect (QHE) where it was found that the edge current was quantized \cite{klitzing, laughlin, thou}.  Investigations related to the existence of edge states and  the geometry of eigenspaces of Schr\"{o}dinger operators has also led to considerable interesting research  \cite{simon,bohm, hatsugai, kane, niu, zak}.  Support for the possible existence of linear unidirectional modes in optical honeycomb lattices was provided in \cite{haldane1, haldane2}. These unidirectional  modes were found to be related to symmetry breaking perturbations which separated the Dirac points in the dispersion surface. The modes are a consequence of a nontrivial integer ``topological" charge associated with the separated bands.

Unidirectional electromagnetic edge modes were first found experimentally in the microwave regime  \cite{wang}. These modes were found  on a square lattice which have no associated Dirac points. Recently though, for photonic graphene, it was shown in \cite{rechts3} that by introducing edges and spatially varying waveguides that unidirectional edge wave propagation at optical frequencies occurs. The waveguides play the role of a pseudo-magnetic field, and in certain parameter regimes, the edge waves are found to be nearly immune to backscattering.  The  pseudo-magnetic fields  used in the experiments \cite{rechts3} are created by periodic changes in the index of refraction of the waveguides in the direction of propagation. The variation in the index of refraction has a well defined helicity and thus breaks `time'-reversal symmetry; here the direction of the wave propagation plays the role of  time.

The analytical description begins with  the lattice nonlinear Schr\"{o}dinger (NLS) equation \cite{rechts3} with cubic Kerr contribution
\begin{equation}
i\p_{z}\psi = - \frac{1}{2k_0} \nabla^{2}\psi +\frac{k_0 \Delta n}{n_0} \psi - \gamma \left|\psi \right|^{2} \psi,
\label{LNLSD}
\end{equation}
where $k_0$ is the input wavenumber, $n_0$ is the ambient refractive index, $\Delta n/n_0$, referred to as the potential, is the linear index change relative to $n_0$, and $\gamma$ represents the nonlinear index contribution. The scalar field $\psi$ is the complex envelope of the electric field, $z$ is the direction of propagation and takes on the role of time, $(x,y)$ is the transverse plane, and $\nabla\equiv(\partial_x,\partial_y)$. Below, in Section~\ref{sec:length-scale}, concrete values are given for the parameters in Eq.~(\ref{LNLSD}).  $\Delta n$ is taken to be a 2D lattice potential in the $(x,y)$-plane which has a prescribed path in the $z$-direction. This path is characterized by a function ${\bf a}(z)=(a_1(z),a_2(z))$, such that after the coordinate transformation
\[
x'=x-a_1(z), ~ y'=y-a_2(z), ~z'=z,
\]
the transformed potential $\Delta n=\Delta n(x',y')$ is independent of $z'$.

Experimentally, the path represented by ${\bf a}(z)$ can be written into the optical material (\eg fused silica) \cite{rechts3} via the femtosecond laser writing technique~\cite{szameit2010}. 
Since this technique enables waveguides to be written along general paths, we only require ${\bf a}(z)$ to be a smooth function. Introducing a transformed field
\[
\psi=\tilde{\psi}\exp\left[\frac{i}{2 k_0} \int_0^z |{\bf A}(\xi)|^2 d\xi\right],
\]
where ${\bf A}$ is induced by the path function ${\bf a}$ via the formula
\begin{equation}
{\bf A}(z)=-k_0{\bf a}'(z),
\label{helicalpseudo}
\end{equation}
Eq.~\eqref{LNLSD} is transformed to
\begin{equation}
i\p_{z'}\tilde{\psi} = - \frac{1}{2k_0} (\nabla' + i {\bf A}(z'))^{2} \tilde{\psi} +\frac{k_0 \Delta n}{n_0} \tilde{\psi} - \gamma \left|\tilde{\psi}\right|^{2} \tilde{\psi}.
\label{MLNLSD}
\end{equation}
In Eq.~(\ref{MLNLSD}),  ${\bf A}$ appears in the same way as if one had added a magnetic field to Eq.~\eqref{LNLSD}; hence  {\bf A} is referred to as a pseudo-magnetic field.

Taking $l$ to be a typical lattice scale size, employing the dimensionless coordinates $x' = lx$, $y'=ly$, $z'=z_*z, z_*=2k_0l^2$, $\tilde{\psi}=\sqrt{P_*} \psi'$, where $P_*$ is input peak power, defining $V(\br')=2k_0^2l^2\Delta n/n_0$ with $\br'\equiv(x',y')$, rescaling ${\bf A}$ accordingly, and dropping the primes, we get the following normalized lattice NLS equation
\begin{equation}
i\p_{z}\psi = -(\nabla + i {\bf A}(z))^{2}\psi +V(\br)\psi - \sigma_0 \left|\psi \right|^{2} \psi.
\label{LNLS}
\end{equation}
The potential $V(\br)$ is taken to be of honeycomb (HC) type.  The dimensionless coefficient $\sigma_0=2\gamma k_0l^2P_*$ is the strength of the nonlinear change in the index of refraction.  We also note that after dropping primes, the dimensionless variables  $x,y,z,\psi$ are used; these dimensionless variables should not be confused with the dimensional variables in Eq. (\ref{LNLSD}).

In contrast to \cite{ACM14fast}, which in turn was motivated by the experiments in \cite{rechts3}, this paper examines the case of periodic pseudo-fields which vary adiabatically, or slowly, throughout the photonic graphene lattice.  We develop an asymptotic theory which leads to explicit formulas for isolated curves in the dispersion relation describing how the structure of edge modes depends on a given pseudo-field ${\bf A}(z)$. Therefore, we can theoretically predict for general pseudo-fields when unidirectional traveling waves exist and the speed with which they propagate.

To exemplify the different classes of dispersion relations allowed in our problem, we take the pseudo-magnetic field to be the following function (``Lissajous'' curves)  \begin{equation}\label{eq:A-lsj}
{\bf A}(z)=(A_1(z),A_2(z))=(\kappa\sin{(D_1\Omega z)},\lambda\sin{(D_2\Omega z+\phi)}),
\end{equation}
where $\kappa$, $\lambda$, $\Omega$, $D_j$, $j=1,2$, and $\phi$ are constant. Below we will consider two cases in detail. In the first case we choose
$D_1=1$, $D_2=1$, $\phi=\pi/2$, and $\kappa=-\lambda$. This corresponds to the pseudo-field employed in \cite{rechts3}, and in most parts of this paper. In this case the above function 
becomes a perfect circle given by
\begin{equation}\label{eq:A-rechts}
{\bf A}(z)= (A_1(z),A_2(z))=\kappa (\sin{\Omega z}, -\cos{\Omega z}),
\end{equation}
where $\kappa$ and $\Omega$ are constant. In the second case we choose $D_1=2$, $D_2=1$, $\phi=\pi/2$, and $\kappa=\lambda$. In this case the above Lissajous curve becomes a figure-8 curve given by
\begin{equation}\label{eq:A-figure-8}
{\bf A}(z)=(A_1(z),A_2(z))=\kappa(\sin{(2\Omega z)},\cos{(\Omega z)}),
\end{equation}
where $\kappa$ and $\Omega$ are constant.  For these classes of pseudo-fields, the numerically computed dispersion relations and the asymptotic prediction of the isolated curves agree very well.  However the dispersion relations also exhibit sensitive behavior in which small gaps in the spectrum appear when the small parameter $\epsilon$ characterizing the slow evolution increases in size.


Further, we find that when edge modes exist there are two important cases.  The first is the case where pure edge modes exist in the entire periodic interval.  The second case is where quasi-edge modes persist only for part of the period. Quasi-edge modes do not exist in the rapidly varying case studied in \cite{ACM14fast}, and so we have shown adiabatic variation of the pseudo-field allows for new dynamics even at the linear level.  We also present  potential scaling regimes where these cases might be observed experimentally; see Section~\ref{sec:length-scale}.

We are also able to analyze  the effect of nonlinearity on these slowly varying traveling edge modes.  A nonconstant coefficient (`time'-dependent) one-dimensional nonlinear Schr\"odinger (NLS) equation governing the envelope of the edge modes is derived and is found to be an effective description of nonlinear traveling edge modes. In the rapidly varying case \cite{ACM14fast}, the associated NLS equation had constant coefficients, and so we see adiabatic variation introduces new dynamics into the nonlinear evolution of edge modes.

Using this new NLS equation, in the focusing case, we find analytically, and confirm numerically, that unidirectionally propagating edge solitons are present in nonlinear photonic graphene lattices. Computation of the NLS equation is compared with direct simulation of the coupled discrete tight binding model with very good agreement obtained. As with the traditional, constant coefficient, focusing NLS equation, nonlinearity balances dispersion to produce nonlinear edge solitons. Depending on the choice of parameters, some of the nonlinear modes appear to be immune to backscattering and can propagate for long distances.  We emphasize that these length scales are far beyond what might be expected from the scales defining the asymptotic theory.  Given this persistence, we argue that such cases are nonlinear analogues of  topologically protected states in topological insulators. Prior to the derivation of the `time' varying NLS equation in this paper, and the constant coefficient version in \cite{ACM14fast}, topological insulators have only been defined for linear systems.  The results in this and previous papers show that nonlinear photonic graphene can also be thought of as a topological insulator.

Time-dependent  NLS equations  also arise when dispersion varies along the propagation direction in an optical fiber \cite{MJA2011,Agrawal}, i.e. the case studied here would be an analogue of  `slow' dispersion management.  Therefore, the results of this paper show that nonlinear, adiabatically varying photonic graphene lattices could provide useful new means for the control of light.  This control results from the merging of nonlinear and symmetry-breaking effects. See also \cite{rechts5}, where bulk nonlinear modes in photonic graphene have been found.  Further, the ``topologically protected nonlinear states" found here can potentially apply to other systems, \eg  recently introduced one dimensional domain walls \cite{FW2014}.
\section{Discrete Equations}\label{discrete eq}
To begin the analysis, the substitution $\psi = e^{-i\br \cdot {\bf A}(z)}\phi$ in $\eq$ \eqref{LNLS} gives
\begin{equation}\label{eq:2d-nls-new}
i\p_{z}\phi = -\Delta \phi -  \br \cdot {\bf A}_{z} \phi + V(\br)\phi- \sigma_0|\phi^{2}|\phi.
\end{equation}
The tight binding approximation for large $V$ assumes a Bloch wave envelope of the form  \cite{abl1}
\begin{equation}\label{eq:tb-ansatz}
\phi \sim \sum_{\bv }\left(a_{\bv}(z)\phi_{1,\bv} + b_{\bv}(z)\phi_{2,\bv} \right)e^{i\bk \cdot \bv}
\end{equation}
where $\phi_{1,\bv}=\phi_{1}(\br-\bv)$, $\phi_{2,\bv}=\phi_{2}(\br-\bv)$ are the linearly independent orbitals associated with the two sites A and B where the honeycomb potential $V(\br)$ has minima in each fundamental cell, and $\bk$ is a vector in the Brillouin zone.  Each $\bv=m\bv_{1}+n\bv_{2}$, where the period vectors $\bv_{1}$ and $\bv_{2}$ are given by
\[
\bv_{1} = \left( \sqrt{3}/2, ~ 1/2\right), ~ \bv_{2} = \left(\sqrt{3}/2, ~ -1/2\right).
\]
Substituting the tight binding approximation (\ref{eq:tb-ansatz}) into Eq.~(\ref{eq:2d-nls-new}), carrying out the requisite calculations (see \cite{abl1} for more details), and after dropping small terms and renormalizing, we arrive at the following two dimensional discrete system
\begin{align}
i\p_{z}a_{mn}  + e^{i\bdd\cdot{\bf A}}(\mathcal{L}_{-}(z)b)_{mn} + \sigma|a_{mn}|^{2}a_{mn}= 0, \label{eq:abmn-a}\\
i\p_{z}b_{mn}  + e^{-i\bdd\cdot{\bf A}}(\mathcal{L}_{+}(z)a)_{mn}  + \sigma|b_{mn}|^{2}b_{mn}= 0, \label{eq:abmn-b}
\end{align}
where
\begin{align*}
(\mathcal{L}_{-}b)_{mn} = & b_{mn} +\rho(b_{m-1,n-1}e^{-i\theta_{1}} +b_{m+1,n-1}e^{-i\theta_{2}}), \\
(\mathcal{L}_{+}a)_{mn} =  & a_{mn} +\rho(a_{m+1,n+1}e^{i\theta_{1}} +a_{m-1,n+1}e^{i\theta_{2}}),
\end{align*}
$\bdd$ is the vector distance between the initial $B$ and $A$ sites (see \cite{ACM14fast}), $\rho$ is a lattice deformation parameter,  $\theta_{1}(z) =  \bv_{1}\cdot(\bk + {\bf A}(z))$, $\theta_{2}(z) = \bv_{2}\cdot(\bk + {\bf A}(z))$, and $\sigma$ is a constant which depends on $\sigma_0$ and the underlying orbitals.  Taking a discrete Fourier transform in $m$, i.e. letting $a_{mn}=a_{n}e^{im\omega}$ and $b_{mn}=b_{n}e^{im\omega}$,  yields the simplified system
\begin{align}
i\p_{z}a_{n}  + e^{i\bdd\cdot{\bf A}}\mathcal{L^-}b_n+\sigma|a_{n}|^{2}a_{n}= 0,\label{eq:abn-ori-a} \\
i\p_{z}b_{n}  +  e^{-i\bdd\cdot{\bf A}}\mathcal{L^+}a_n +\sigma|b_{n}|^{2}b_{n}= 0,\label{eq:abn-ori-b}
\end{align}
where
\begin{align}
\mathcal{L^-}b_n= \left( b_{n} +\rho\gamma^{\ast}(z;\omega)b_{n-1}\right)\\
\mathcal{L^+}a_n=\left(a_{n} +\rho\gamma(z;\omega)a_{n+1}\right)
\end{align}
$\gamma(z;\omega) = 2e^{i\varphi_{+}(z)}\cos(\varphi_{-}(z)-\omega)$ and
\[
\varphi_{+}(z) = (\theta_{2}(z)+\theta_{1}(z))/2, ~ \varphi_{-}(z) = (\theta_{2}(z)-\theta_{1}(z))/2.
\]

Since $\gamma(z;\omega+\pi)=-\gamma(z;\omega)$, if $(a_n(z),b_n(z))$ is a solution of Eqs.~(\ref{eq:abn-ori-a}--\ref{eq:abn-ori-b}) at any given $\omega$, then $(-1)^n(a_n(z),b_n(z))$ is a solution of Eqs.~(\ref{eq:abn-ori-a}--\ref{eq:abn-ori-b}) at $\omega+\pi$. Therefore, in the following $\omega$ is taken to be defined on a periodic interval of length $\pi$. The value of $\bk$ is be taken to be zero; we assume that ${\bf A}(z)$ has zero mean (see \cite{ACM14fast}).

To analyze the equations (\ref{eq:abn-ori-a}-\ref{eq:abn-ori-b}) we will work on a zig-zag edge, which in the semi-infinite case means letting $a_{n}(z)=0, n\leq 0$. We further assume that the pseudo-field ${\bf A}$ evolves adiabatically, i.e.  we take $Z=\epsilon z$ and   ${\bf A}= {\bf A}(Z)$. Our goal is to find slowly evolving and decaying modes as $n \rightarrow \infty$. It is convenient to employ a multiple-scales ansatz
\[
a_{n} = a_{n}(z,Z), ~ b_{n}= b_{n}(z,Z),
\]
where $Z = \epsilon z$.  From Eq. (\ref{eq:abn-ori-a}-\ref{eq:abn-ori-b}) the basic perturbation equations are given by
\begin{align}
i\p_{z}a_{n}  + e^{i\bdd\cdot{\bf A}}\mathcal{L^-}b_n=  -\epsilon \left( ia_{n,Z}+\tilde{\sigma} |a_{n}|^{2}a_{n}\right), \label{eq:abn1-ori-a} \\
i\p_{z}b_{n}  +  e^{-i\bdd\cdot{\bf A}}\mathcal{L^+}a_n=   -\epsilon \left( ib_{n,Z}+\tilde{\sigma} |b_{n}|^{2}b_{n}\right), \label{eq:abn1-ori-b}
\end{align}
where $\sigma= \epsilon\tilde{\sigma}$. Then expanding $a_{n},b_{n}$ in powers of $\epsilon$
\[
a_{n} = a_{n}^{(0)} + \epsilon a_{n}^{(1)} + \cdots, ~ b_{n} = b_{n}^{(0)} + \epsilon b_{n}^{(1)} + \cdots,
\]
we have at leading order
\begin{align}
i\p_{z}a^{(0)}_{n}  + e^{i\bdd\cdot{\bf A}}\mathcal{L^-}b^{(0)}_n=  0 \\
i\p_{z}b^{(0)}_{n}  +  e^{-i\bdd\cdot{\bf A}}\mathcal{L^+}a^{(0)}_n=   0
\end{align}
which has  the following edge solution \cite{abl5}
\begin{equation}
b_{n}^{(0)}(Z) = C(Z,\omega)b_{n}^S(Z), ~ a_{n}^{(0)} = 0,
\label{envelope_eq}
\end{equation}
where
\begin{equation}
b_{n}^S(Z) = (1-\rho^2|\gamma(Z)|^2)^{1/2}\left(-\rho \gamma^{\ast}(Z) \right)^{n},
\label{statsoln}
\end{equation}
with $b_{n}^S=0, n < 0$.
This generalizes \cite{abl5} where purely stationary edge modes are obtained. The function $C(Z,\omega)$ is determined below.
\section{Linear system}\label{4}
In the linear problem we take $\tilde{\sigma}=0$. We first return to investigate the properties of the edge solution (\ref{envelope_eq}-\ref{statsoln}).
In order to have decaying modes ($b_{n}^S \rightarrow 0$  as  $n \rightarrow \infty$)  we need $\rho|\gamma(\bar{Z};\omega)|<1$, which then requires us to take
\begin{equation}\label{eq:omega-varphi-theta}
\omega -\varphi_{-}(Z) \in \left\{\begin{array}{cc}
\left(\tilde{\theta},\pi-\tilde{\theta}\right), & \rho\geq1/2;\\
S^1, & \rho<1/2,
\end{array}\right.
\end{equation}
where $\tilde{\theta}=\cos^{-1}(1/(2\rho))$ and $S^1\equiv\mathbb{R}/(\pi\mathbb{Z})$.

Fixing the frequency $\omega$, we define the time interval $\mathcal{I}_{Z}(\omega)$ to be
\[
\mathcal{I}_{Z}(\omega) = \left\{Z : \mbox{Eq.}~\eqref{eq:omega-varphi-theta}~\mbox{is satisfied}  \right\}.
\]
If $\mathcal{I}_Z(\omega)=[0,T]$, then the edge mode remains localized for the entire period for the given frequency $\omega$. These edge modes will be referred to as pure edge modes. If $\emptyset\subset\mathcal{I}_Z(\omega)\subset[0,T]$, then the edge mode remains localized for only part of the period after which the mode disintegrates into the bulk. These edge modes are referred to as quasi-edge modes. In applications, pure edge modes are expected to be more relevant since they can in principle propagate over many periods.  We define $\mathcal{I}_{p}$ to be the frequencies $\omega$ for which pure edge modes exist, i.e.
\[
\mathcal{I}_p= \left\{ \omega : \mathcal{I}_{Z}(\omega) = [0,T] \right\}.
\]
In a single period, $\varphi_-(Z)$ spans an interval denoted by $[\varphi_-^\textrm{min},\varphi_-^\textrm{max}]$. The frequency interval $\mathcal{I}_p$ of pure edge modes then falls into one of the following three cases
\begin{align*}
\textrm{Case (I): }   \mathcal{I}_p=&S^{1},\,\rho<1/2;\\
\textrm{Case (II): }  \mathcal{I}_p=&\left(\omega_-,\omega_+\right),\,\rho\geq1/2 \textrm{ and } \omega_-<\omega_+,\\
\textrm{Case (III): } \mathcal{I}_p=&\emptyset,\,\rho\geq1/2 \textrm{ and } \omega_-\geq\omega_+,
\end{align*}
where $\omega_-\equiv\varphi_-^\textrm{max}+\tilde{\theta}$ and $\omega_+\equiv\varphi_-^\textrm{min}+\pi-\tilde{\theta}$.  We note that since no pure edge mode exists for any range of frequencies $\omega$ in Case (III), this case will be omitted in the following discussion of pure edge modes. 

Next we note that associated with the edge solution (\ref{envelope_eq}-\ref{statsoln}), the Fredholm condition (\ref{Fredholm}), derived in Appendix~\ref{pert}, implies that the envelope of the edge solution satisfies the equation
\[
\p_ZC+i\alpha_l(Z;\omega)C=0,
\]
where
\begin{equation}
\alpha_l(Z;\omega)=-i\ipro{\p_{Z}b^{S}}{b^{S}} = -i\sum_{l=0}^{\infty}\p_{Z}b_{l}^S(Z)(b_l^S)^{\ast}(Z).
\label{alphaterms}
\end{equation}
Substituting (\ref{statsoln}) into \eqref{alphaterms} and manipulating terms, we find that
\begin{align*}
\alpha_l(Z;\omega)= & -i\frac{\rho^{2}}{2(1-\rho^{2}|\gamma|^{2})}\left(\gamma \p_{Z}\gamma^{\ast} - \gamma^{\ast}\p_{Z}\gamma \right) \\
= & -4\rho^{2}\frac{\p_{Z}(\varphi_{+})\cos^{2}\left(\varphi_{-}(Z)-\omega \right)}{1-4\rho^{2}\cos^{2}\left(\varphi_{-}(Z)-\omega\right)}.
\end{align*}
As can be readily seen, this quantity is real, and thus since
\[
C(Z;\omega) = C(0) \mbox{exp}\left(-i\int^{Z}_{0} \alpha_l(t;\omega) dt \right) ,
\]
this term shows that the influence of the nontrivial pseudo-magnetic field ${\bf A}$ on the nearly stationary edge modes is the introduction of a non-trivial phase.  We know that the function $\alpha_l(Z)$ is periodic in $Z$, and so we can write
\[
 \alpha_l(Z;\omega) = \bar{\alpha}(\omega)  + \sum_{k\neq 0} \hat{\alpha}_{k}(\omega)e^{i2\pi kZ/T},
\]
where the average term $\bar{\alpha}(\omega)$ is given by
\begin{equation}\label{eq:talpha-omega}
\bar{\alpha}(\omega) = \frac{1}{T}\int_{0}^{T} \alpha_l(t;\omega) dt.
\end{equation}

Therefore, to leading order, we have that the edge mode in the presence of a slowly varying pseudo-field ${\bf A}(Z)$ is given by
\begin{equation}
b^{(0)} = e^{-i \bar{\alpha}(\omega) Z} p(Z)b^{S}(Z),
\label{zeromode}
\end{equation}
where the periodic function $p(Z)$ is given by
\[
p(Z) = \mbox{exp}\left(-\frac{T}{2\pi}\sum_{k\neq 0} \frac{\hat{\alpha}_{k}}{k}\left(e^{i2\pi k Z/T}-1 \right) \right).
\]
Since both $p(Z)$ and $b^{S}(Z)$ are periodic in $Z$, we see the Floquet parameter for the periodic problem is given by $\bar{\alpha}(\omega)$, which then gives us the dispersion relation in the presence of a nontrivial pseudo-field ${\bf A}$. In general, one cannot obtain Floquet parameters in explicit form, though in this perturbative system we can do so.

The dispersion relation $ \bar{\alpha} (\omega)$ can be classified according to a $\mathbb{Z}_2$ topological index $I\equiv N(\mathrm{mod}\,2)$ where $N$ is the number of roots of $\bar{\alpha}(\omega)=0$~\cite{ACM14fast}.
It can be readily seen that $I=0$ in Case (I), or when $\mathcal{I}_p=S^1$.
In Case (II), when $\mathcal{I}_p=(\omega_-,\omega_+)$, $ \bar{\alpha} (\omega)$ has the asymptotic behavior for $\omega_{-} < \omega < \omega_{+}$ (see Appendix~\ref{app:alpha-asym})
\begin{align}
\bar{\alpha}(\omega\rightarrow \omega_{-})&=-C_{l}A_1'(Z_-)(\omega-\omega_-)^{-1/2},\label{eq:alpha-asym-m}\\
\bar{\alpha}(\omega\rightarrow \omega_{+})&=-C_{r}A_1'(Z_+)(\omega_+-\omega)^{-1/2},\label{eq:alpha-asym-p}
\end{align}
where $C_{l/r}$ are positive constants, and $Z_\pm$ are those times such that
\[
\varphi_{-}\left( Z_{-}\right) = \varphi_{-}^{\max}, ~ \varphi_{-}\left( Z_{+}\right) = \varphi_{-}^{\min}.
\]
Therefore the direction of blowup of $ \bar{\alpha} (\omega)$ as $\omega\rightarrow\omega_\pm$ agrees with the sign of $-A_1'(Z)$ at $Z=Z_\pm$. 
It follows that for a pseudo-field ${\bf A}(Z)$ which is counterclockwise in the $(A_1,A_2)$-plane, $ \bar{\alpha} (\omega)$ goes from $-\infty$ at $\omega=\omega_-$ to $+\infty$ at $\omega_+$, and vice-versa for a clockwise pseudo-field. Thus for any pseudo-field ${\bf A}(Z)$ that forms a simple closed curve in the $(A_1,A_2)$-plane, the topological index is always $I=1$ in Case (II), and the sign of the overall group velocity depends only on the helicity of the pseudo-field. However, if the pseudo-field forms a self-intersecting closed curve, then $\bar{\alpha}(\omega)$ may asymptote to $-\infty$ or $+\infty$ at both $\omega=\omega_-$ and $\omega=\omega_+$, in which case the topological index is $I=0$.

In the case where the topological index is nontrivial, i.e.~$I=1$, the pure edge modes are expected to behave like topologically protected modes, i.e.~there are no backward propagating modes to inhibit the evolution. Below we show numerical examples of such long lasting edge states. A detailed discussion of the mechanism of topological protection is outside of the scope of this paper.

\begin{figure}
\centering \includegraphics[width=0.28\textwidth]{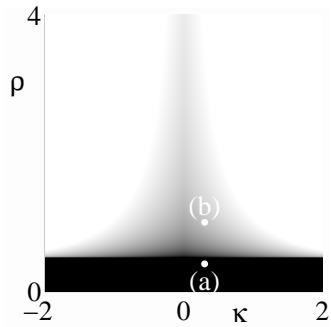}
\caption{The width $|\mathcal{I}_p|$ of the existence interval $\mathcal{I}_p$ of pure edge states as a function of the deformation parameter $\rho$ and the pseudo-field parameter $\kappa$. The color scale interpolates between white for $0$ and black for $\pi$. The dispersion relations at the labeled points (a)--(b) are shown in Fig.~\ref{fig:EDGE_disp_pure}.
}
\label{fig:EDGE_slow_twopar}
\end{figure}

To make the above analysis more concrete, we take the pseudo-field ${\bf A}(Z)$ to be Eq.~(\ref{eq:A-rechts}) with $\Omega=\epsilon$, unless otherwise stated. As shown in Fig.~\ref{fig:EDGE_slow_field}(a) for $\kappa=0.3$, ${\bf A}(Z)$ forms a counterclockwise circle in the $(A_1,A_2)$-plane.  In this case $\mathcal{I}_p$ is always centered around $\pi/2$ since $\varphi_-^\textrm{min}+\varphi_-^\textrm{max}=0$. Figure~\ref{fig:EDGE_slow_twopar} shows the width $|\mathcal{I}_p|$ of $\mathcal{I}_p$ as a function of $\rho$ and $\kappa$. For $\rho<1/2$, $|\mathcal{I}_p|$ is always $\pi$. As $\rho$ increases past $1/2$, $|\mathcal{I}_p|$ decreases discontinuously as a function of $\rho$ for $\kappa\neq0$ with the discontinuity increasing as $|\kappa|$ increases. For $\rho>1/2$, $|\mathcal{I}_p|$ decreases as $\rho$ or $|\kappa|$ increases and becomes $0$ for sufficiently large $\rho$ or $|\kappa|$.

\begin{figure}
\centering
\begin{tabular}{ccc}
\includegraphics[width=.32\textwidth]{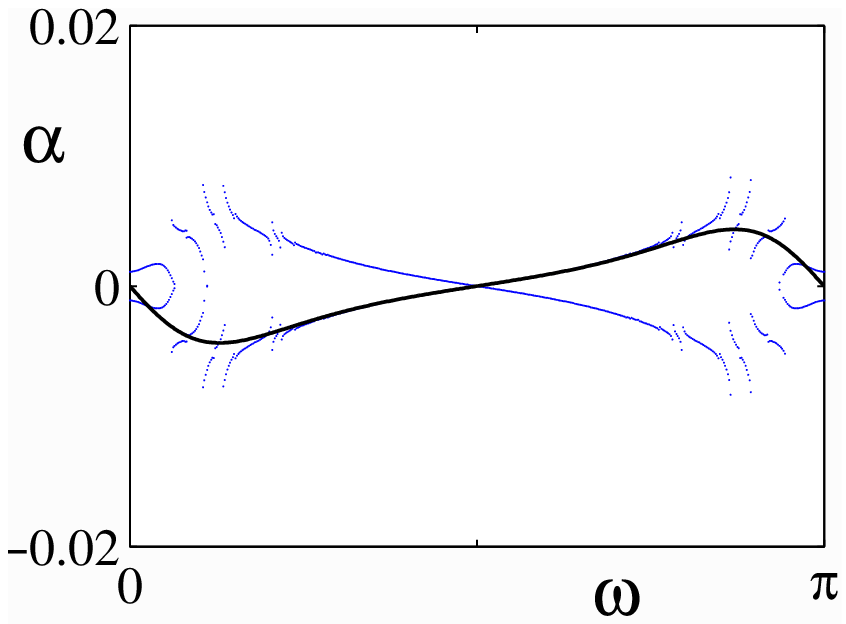} &
\includegraphics[width=.32\textwidth]{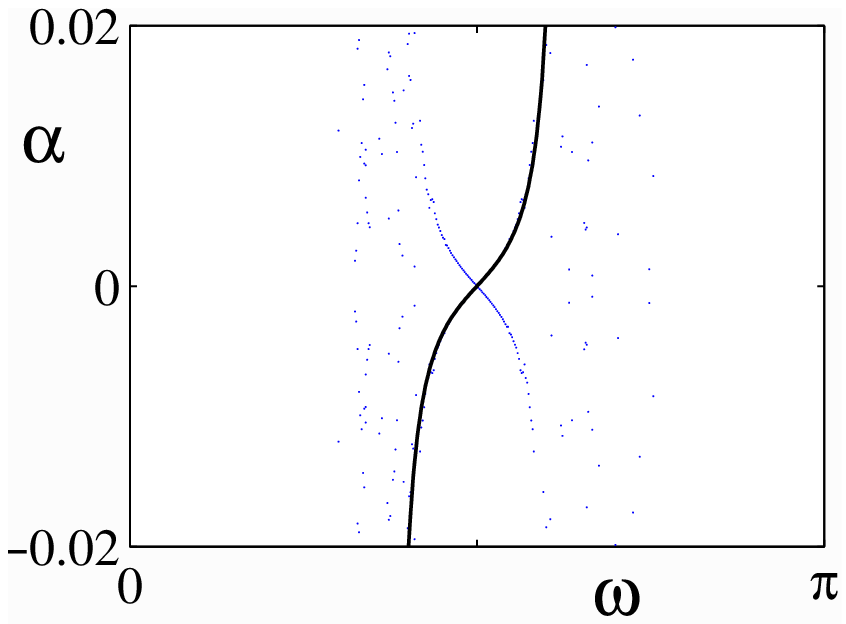} &
\includegraphics[width=.32\textwidth]{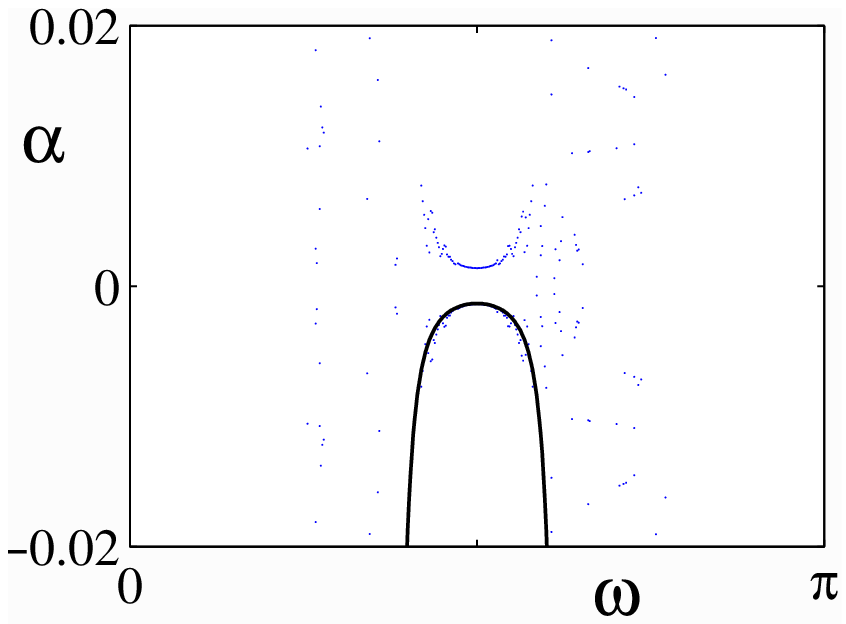} \\
(a) & (b) & (c)
\end{tabular}
\caption{The dispersion relation of pure edge modes for points (a) and (b) in Fig.~\ref{fig:EDGE_slow_twopar} computed using the circular pseudo-field Eq.~(\ref{eq:A-rechts}) with $\kappa=0.3$ and (a) $\rho=0.4$; (b) $\rho=1$. Panel (c) is computed using the Lissajous pseudo-field Eq.~(\ref{eq:A-lsj}) with $\kappa=\lambda=0.3$, $D_1=2$, $D_2=1$, $\phi=\pi/2$, and $\rho=1$. The number of lattice sites is $40$ and the small parameter is $\epsilon=2\pi/60$. The blue curve represents pure edge modes computed numerically from Eqs.~(\ref{eq:abn-ori-a}--\ref{eq:abn-ori-b}). The black curve shows the asymptotic prediction Eq.~(\ref{eq:talpha-omega}).}
\label{fig:EDGE_disp_pure}
\end{figure}

\begin{figure}
\centering
\begin{tabular}{cc}
\includegraphics[width=.16\textwidth]{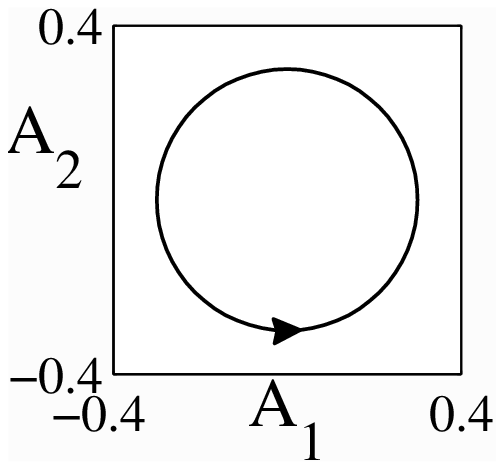} &
\includegraphics[width=.16\textwidth]{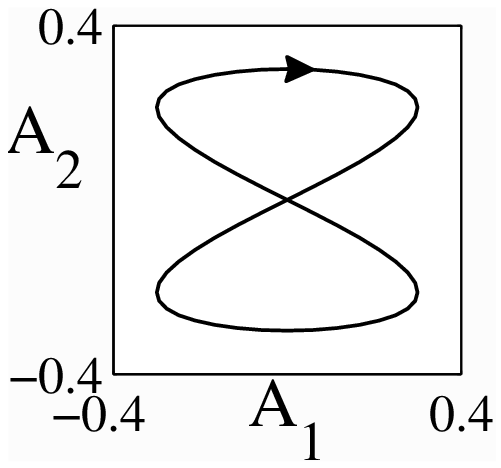} \\
(a) & (b)
\end{tabular}
\caption{Plots of the pseudo-field ${\bf A}(Z)=(A_1(Z),A_2(Z))$ corresponding to the parameters used in (a) Fig.~\ref{fig:EDGE_disp_pure}(a,b); (b) Fig.~\ref{fig:EDGE_disp_pure}(c).} \label{fig:EDGE_slow_field}
\end{figure}

Figure~\ref{fig:EDGE_disp_pure} shows the unscaled dispersion relation $\alpha(\omega)=\epsilon \bar{\alpha} (\omega)$ of pure edge modes at the labeled points (a)-(b) on the $(\rho,\kappa)$-plane in Fig.~\ref{fig:EDGE_slow_twopar}.  The blue curves show the dispersion relations computed directly using Eqs.~(\ref{eq:abn-ori-a}-\ref{eq:abn-ori-b}).  The computational domain has $40$ lattice sites for each vector $a$ and $b$ with zig-zag boundary conditions on both ends.   The black curve shows the asymptotically predicted dispersion relation Eq.~(\ref{eq:talpha-omega}) for pure edge modes localized on the left. In Fig.~\ref{fig:EDGE_disp_pure}(a) and Fig.~\ref{fig:EDGE_disp_pure}(b), the small parameter is chosen to be $\epsilon=2\pi/60$. The asymptotic theory agrees well with the numerical computation. As $\epsilon$ decreases, the theory improves further. Fig.~\ref{fig:EDGE_disp_pure}(a) shows a Case (I) dispersion relation computed at $\rho=0.4$ and $\kappa=0.3$. In this case $\alpha(\omega)$ crosses the $\omega$ axis twice, so the topological index is $I=0$. Fig.~\ref{fig:EDGE_disp_pure}(b) shows a Case (II) dispersion relation computed at $\rho=1$ and $\kappa=0.3$. In this case, $\alpha(\omega)$ crosses the $\omega$ axis once, so the topological index is $I=1$.
Interestingly, the dispersion relation of edge modes in Fig.~\ref{fig:EDGE_disp_pure}(a) breaks up into multiple segments near $\omega=0,\pi$.
Also, in Fig.~\ref{fig:EDGE_disp_pure}(b) there are scattered eigenvalues near $\alpha=0$ for $\omega$ immediately outside $\mathcal{I}_p=(\omega_-,\omega_+)$. These eigenvalues may result from the presence of quasi-edge modes at these values of $\omega$.

To illustrate the possibility of an $I=0$ (topologically trivial) dispersion relation in Case (II), we take the pseudo-field ${\bf A}(Z)$ to be Eq.~(\ref{eq:A-figure-8}). As shown in Fig.~\ref{fig:EDGE_slow_field}(b) for $\kappa=0.3$, ${\bf A}(Z)$ forms a figure-8 curve in the $(A_1,A_2)$-plane. Since $A_1'(Z_\pm)>0$ at $Z=Z_{\pm}$
Eqs.~(\ref{eq:alpha-asym-m}--\ref{eq:alpha-asym-p}) imply that $\bar{\alpha}(\omega)$ tends to $-\infty$ at both $\omega=\omega_{\pm}$. Fig.~\ref{fig:EDGE_disp_pure}(c) shows the Case (II) dispersion relation computed using this pseudo-field at $\rho=1$. As predicted, $\bar{\alpha}(\omega)$ tends to $-\infty$ at $\omega=\omega_\pm$, so the topological index is $I=0$.

\begin{figure}
\centering \includegraphics[width=0.32\textwidth]{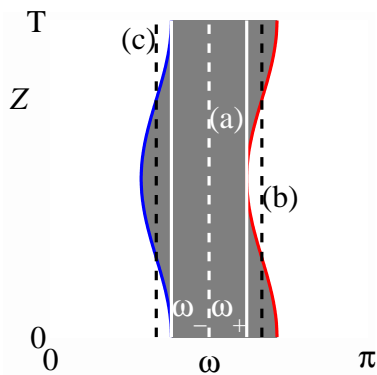}
\caption{The shaded region shows the region in the $(\omega,Z)$-plane determined by the localization criterion Eq.~(\ref{eq:omega-varphi-theta}) at the same parameters as in Fig.~\ref{fig:EDGE_disp_pure}(b). The two solid white lines bound the existence interval $\mathcal{I}_p$ of pure edge modes. The dashed lines show the values of $\omega$ used in the panels of Fig.~\ref{fig:EDGE_slow_exic}.} \label{fig:EDGE_slow_interval}
\end{figure}

\begin{figure}
\centering
\begin{tabular}{ccc}
\includegraphics[width=0.32\textwidth]{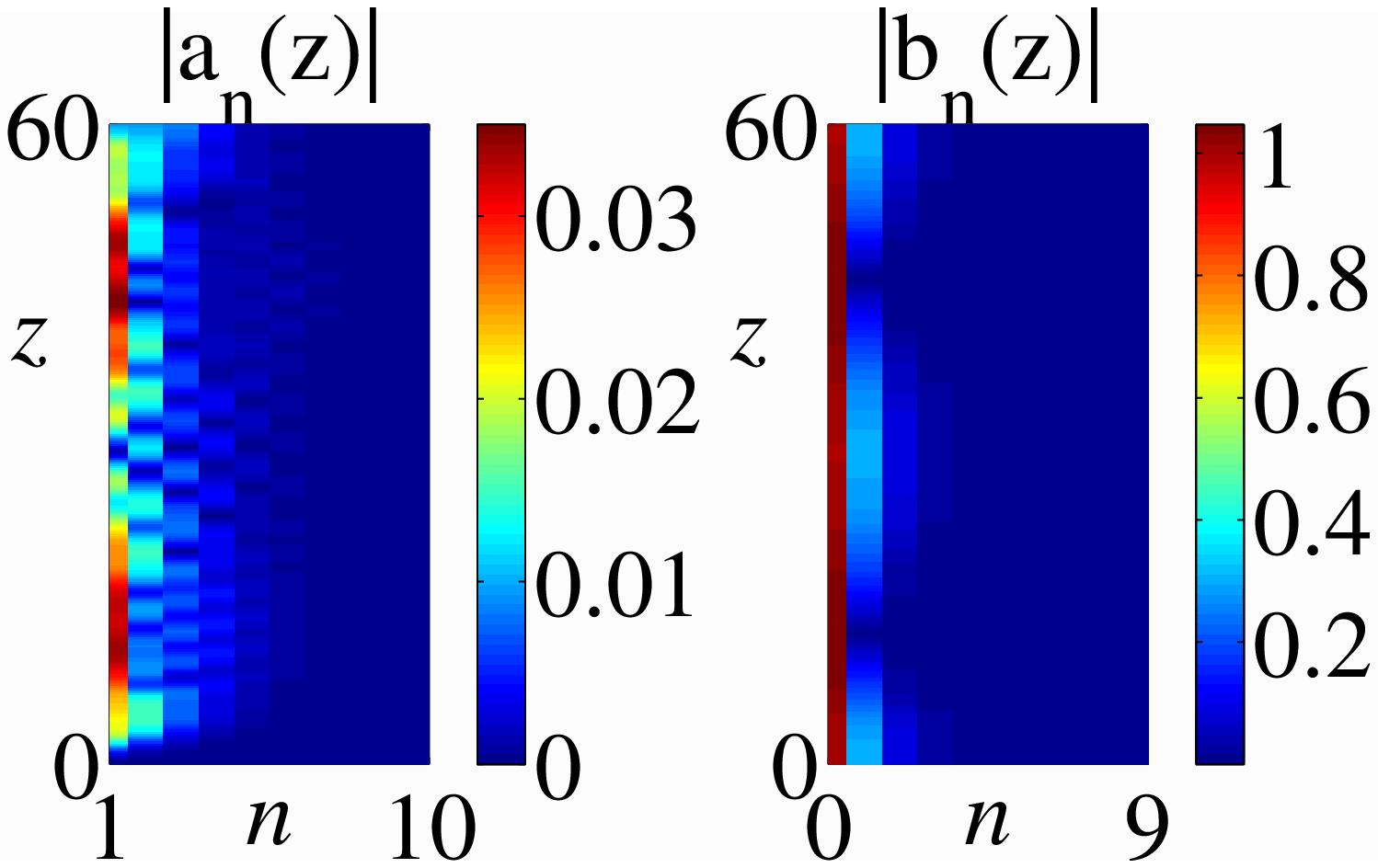} &
\includegraphics[width=0.32\textwidth]{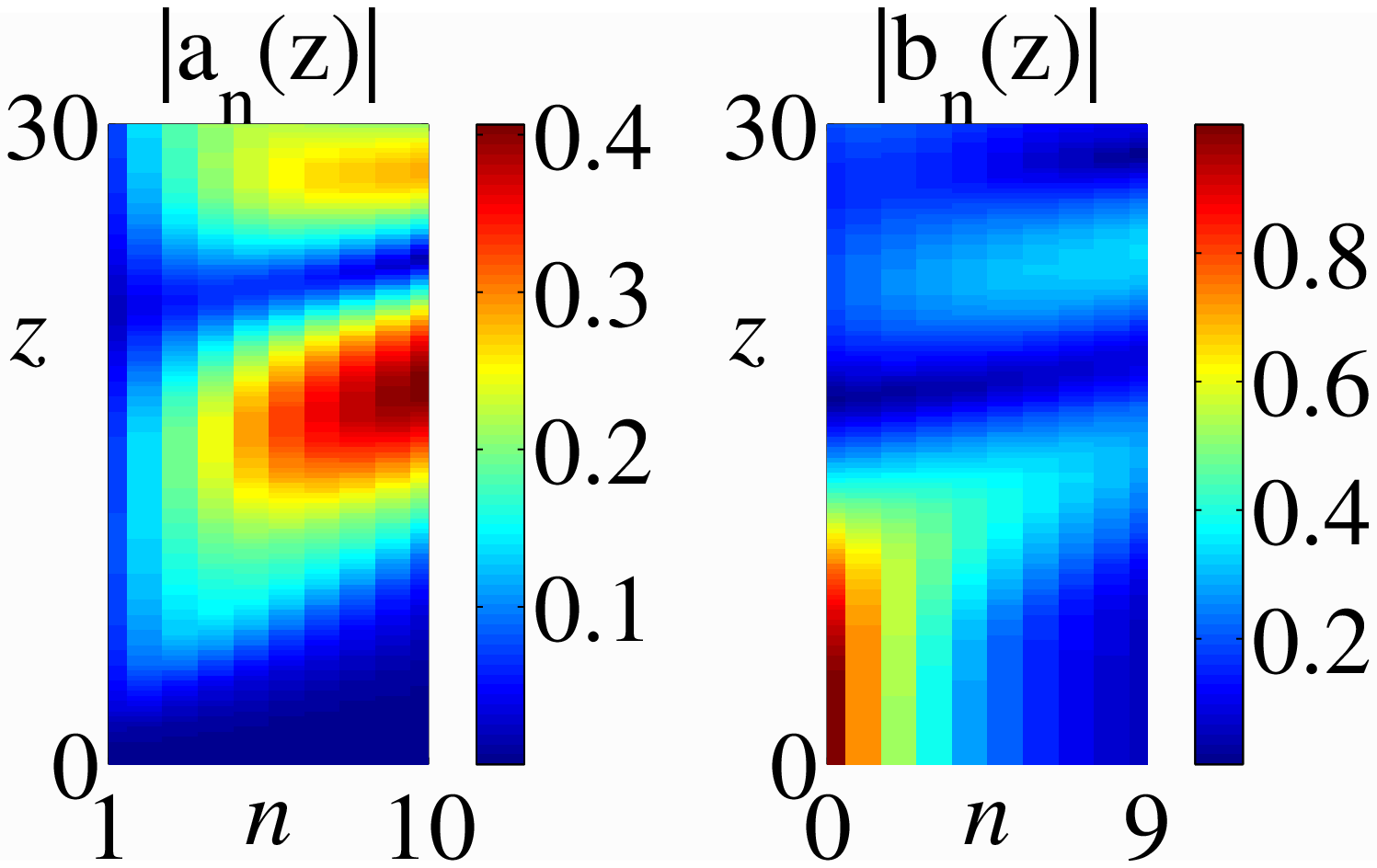} &
\includegraphics[width=0.32\textwidth]{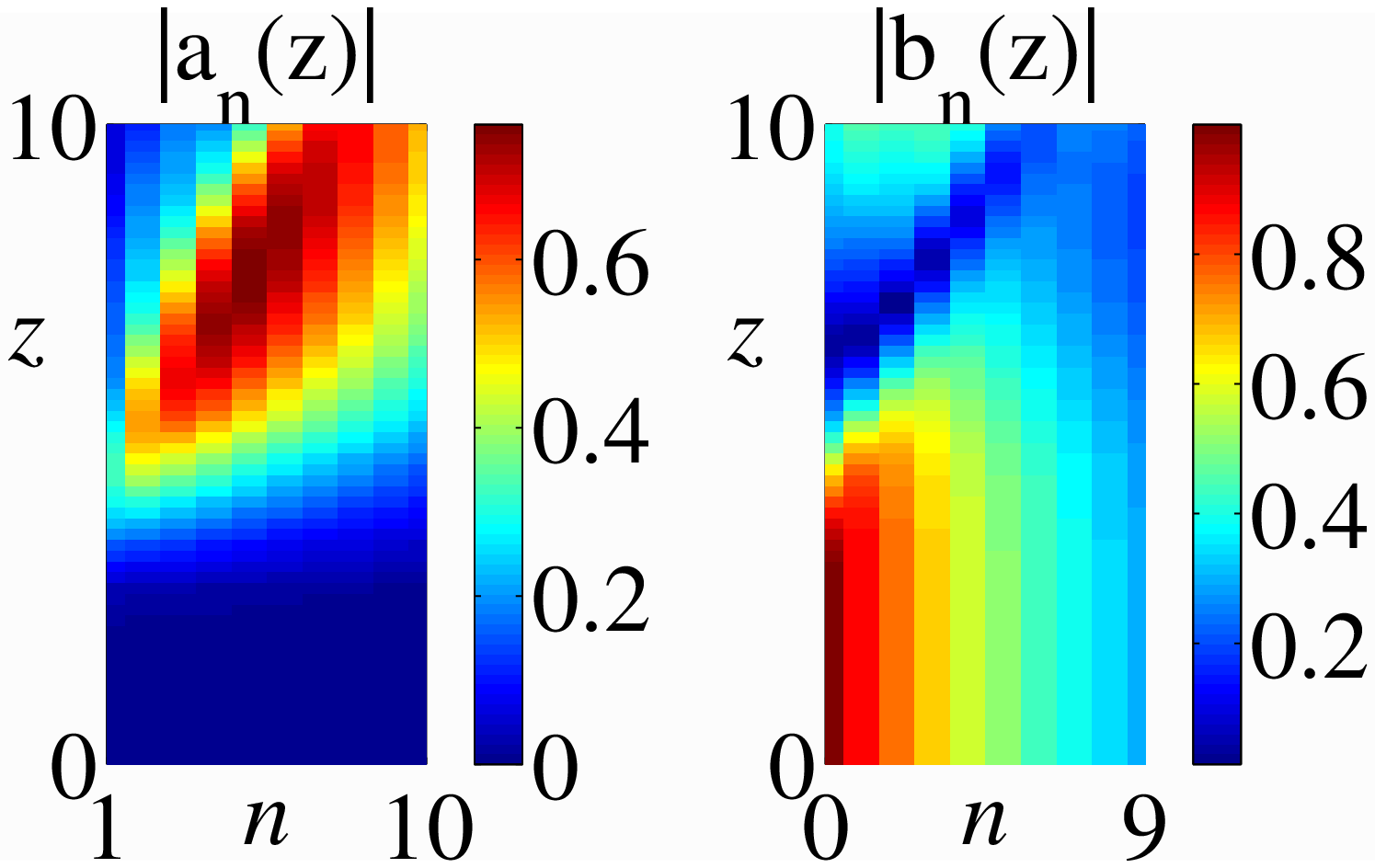} \\
(a) & (b) & (c)
\end{tabular}
\caption{Time evolution of Eq.~(\ref{eq:abn-ori-a}--\ref{eq:abn-ori-b}) with parameters $(\rho,\kappa)=(1,0.3)$ and $\epsilon=2\pi/60$ as in Fig.~\ref{fig:EDGE_disp_pure}(b). The initial condition is given by Eq.~(\ref{eq:bnL}) with $(\omega,\delta)$: (a) $(\pi/2,1)$; (b) $(2\pi/3,1)$; (c) $(\pi/3,0.7)$. The number of lattice sites is $80$, but only the leftmost $10$ sites are shown.}
\label{fig:EDGE_slow_exic}
\end{figure}

Next we return to the Case (II), $I=1$ dispersion relation in Fig.~\ref{fig:EDGE_disp_pure}(b), computed using the circular pseudo-field Eq.~(\ref{eq:A-rechts}) with $\kappa=0.3$. As stated above, this is a topologically nontrivial case. The region in the $(\omega,Z)$-plane determined by the localization criterion Eq.~(\ref{eq:omega-varphi-theta}) is shown as the shaded region in Fig.~\ref{fig:EDGE_slow_interval}. Thus the localization interval $\mathcal{I}_Z(\omega)$ corresponds to the vertical slice through the shaded region at fixed $\omega$, and the existence interval $\mathcal{I}_p=(\omega_-,\omega_+)$ of pure edge modes is bounded by the two solid white lines. The values of $\omega$ used in the panels of Fig.~\ref{fig:EDGE_slow_exic} below are shown as the dashed lines in Fig.~\ref{fig:EDGE_slow_interval}.

Figure~\ref{fig:EDGE_slow_exic} shows the time evolution of Eq.~(\ref{eq:abn-ori-a}--\ref{eq:abn-ori-b}) using the localized initial condition
\begin{equation}\label{eq:bnL}
a_n^L=0,\quad b_n^L(\delta)=(-\rho\delta\gamma^\ast(0))^n,
\end{equation}
where $0<\delta\leq1$. At $\delta=1$, $b_n^L$ becomes the stationary mode $b_n^S$ rescaled such that $b_0^L=1$.  Figure~\ref{fig:EDGE_slow_exic}(a) shows the time evolution at $\omega=\pi/2\in\mathcal{I}_p$ and $\delta=1$. In this case $\mathcal{I}_Z(\omega)=[0,T]$, or a pure edge mode exists for these parameter values. Thus the initial condition (\ref{eq:bnL}) with $\delta=1$ remains localized for the entire period, with most power remaining in $b_n$. Figure~\ref{fig:EDGE_slow_exic}(b) shows the time evolution at $\omega=2\pi/3\not\in\mathcal{I}_p$ and $\delta=1$. In this case $\mathcal{I}_Z(\omega)=[-T/4,T/4]$ is centered around $Z=0$. Thus the initial condition (\ref{eq:bnL}) with $\delta=1$ remains localized for part of the period before disintegrating into the bulk with power distributed into both $a_n$ and $b_n$. Figure~\ref{fig:EDGE_slow_exic}(c) shows the time evolution at $\omega=\pi/3\not\in\mathcal{I}_p$ and $\delta=0.7$. In this case $\mathcal{I}_Z(\omega)=[T/4,3T/4]$ is centered around $Z=T/2$. Thus the initial condition (\ref{eq:bnL}) is no longer localized at $\delta=1$; the artificially constructed localized initial condition with $\delta=0.7$ rapidly disintegrates into the bulk.

We remark that pure edge modes with power concentrated in either the $a$ or $b$ lattice sites (i.e. edge modes localized on the right/left respectively), whose dispersion relations are shown in Fig.~\ref{fig:EDGE_disp_pure}, are not the only localized eigenmodes in the Floquet spectrum computed numerically using Eqs.~(\ref{eq:abn-ori-a}--\ref{eq:abn-ori-b}). Near $\omega=\pi/2$, there are additional localized eigenmodes with power equally distributed in the $a$ and $b$ lattice sites. These eigenmodes are not exponentially localized and thus span many more lattice sites than pure edge modes. Moreover, they span more lattice sites as $\epsilon$ decreases, in contrast to pure edge modes whose decay exponent in $n$ is independent of $\epsilon$. These eigenmodes share some common features with Tamm-like edge states near Van Hove singularities observed in Ref.~\cite{Segev4}.
\subsection{Numerical Computation of Dispersion Relations}
It is interesting to see where the dispersion relations of pure edge modes as shown in Fig.~\ref{fig:EDGE_disp_pure} lie in the full Floquet spectra.  The Floquet spectrum is computed using Eqs.~(\ref{eq:abn-ori-a}--\ref{eq:abn-ori-b}) with a finite number of lattice sites, taken to be $d/2$, for each vector $a$ and $b$ and zig-zag boundary conditions on both ends.  Defining the $d\times d$ $T$-periodic Hermitian matrix
\[
\mathcal{L}(z)=\bp 0 & e^{i\bdd\cdot{\bf A}(z)}\mathcal{L}^{-}(z) \\ e^{-i\bdd\cdot{\bf A}(z)}\mathcal{L}^{+}(z) & 0 \ep,
\]
by using Floquet's theorem \cite{chicone} on the periodic matrix problem
\begin{equation}\label{eq:lin-ode-ori}
i\partial_zQ+\mathcal{L}(z)Q=0, ~ Q(0) = I_{d},
\end{equation}
where $I_{d}$ is the $d\times d$ identity matrix, we have that
\[
Q(z) = V(z) e^{i\Lambda z} V^{-1}(0)
\]
where $\Lambda$ is a $d\times d$ diagonal matrix with real diagonal entries $\lambda_{j}$ and $V(z)$ is a $T$ periodic $d\times d$ matrix.  The values $\lambda_{j}$ are the Floquet spectrum associated with the periodic problem \eqref{eq:lin-ode-ori}, and the columns of $V(z)$, denoted by $V_j(z)$, are the Floquet eigenvectors.  Numerically, $V$ and $\Lambda$ may be computed in the following two steps.

In the first step, we solve Eq.~(\ref{eq:lin-ode-ori}) up to the period $T$. The eigenvector and eigenvalue matrices of the final state $Q(T)$ are then respectively $V(T) = V(0)$ and $e^{i\Lambda T}$, and we have each entry of the matrix $Q(z)$ for $0\leq z \leq T$.  From this, we can 
determine each eigenvalue $\lambda_{j}$ up to an integer multiple of $2\pi/T$; this represents an ambiguity that cannot be resolved by studying $Q(T)$ alone. Hence in the second step, for each Floquet eigenvector $V_{j}(z)$, we consider its time evolution given by $Q(z)V_j(0)$, and define its inner product $p_{j}(z)$ with a time-independent $d$-dimensional vector $f$ as
\begin{equation}
p_{j}(z)\equiv\langle f,Q(z)V_j(0)\rangle =\langle f,V_{j}(z)e^{i\lambda_{j}z}\rangle  ,
\end{equation}
where
\begin{equation}
\langle g,h\rangle\equiv\sum_{l}g_{l}^*h_{l},
\end{equation}
and determine $\lambda_{j}$ via
\begin{equation}
p_{j}(T)= p_{j}(0)e^{i\lambda_{j}T}.
\end{equation}
Using Eq.~(\ref{eq:lin-ode-ori}), the phase of $p_{j}(z)$, defined as
\begin{equation}
\phi_{j}(z)\equiv-i\log{p_{j}(z)},
\end{equation}
evolves as
\begin{equation}
\frac{d\phi_{j}(z)}{dz}= \frac{\langle f,\mathcal{L}(z)Q(z)V_j(0)\rangle}{\langle f,Q(z)V_j(0)\rangle}
= \frac{\langle f,\mathcal{L}(z)V_{j}(z)\rangle}{\langle f,V_{j}(z)\rangle}, \label{eq:pseudo-phase-evo}
\end{equation}
which can then be integrated from $0$ to $T$ to yield
\begin{equation}
\lambda_{j}=\Delta\phi_{j}/T,\quad\Delta\phi_{j}\equiv\phi_{j}(T)-\phi_{j}(0).
\end{equation}
Since $\Delta\phi_{j}$ is $2\pi$ times the winding number of $p_{j}(z)$ around the origin for $z\in[0,T]$, it is unique modulo $2\pi$ and thus $\lambda_j$ is unique modulo $2\pi/T$. We further emphasize that since the computation only relies on $Q(z)$ and $V(0)$, both of which have been computed in the first step, and since no explicit use of a logarithm is made, we have removed any ambiguity in computing the Floquet spectrum.  If the evolution operator $\mathcal{L}$ is independent of $z$, then $V_j(z)=V_j(0)$ and $\mathcal{L}V_{j}(0)=\lambda_{j}V_{j}(0)$, so the above procedure indeed leads to the correct eigenvalue $\lambda_{j}$ independent of $f$, as long as $\langle f,V_{j}(0)\rangle\neq0$.

However if $\mathcal{L}$ depends on $z$, then so does the Floquet eigenvector $V_{j}(z)$, and $\lambda_{j}$ may depend on the choice of $f$. Let us consider the particular choice $f=V_{j}(z_0)$ where $z_0$ is arbitrary. It can be seen that if the correlation function
\begin{equation}\label{eq:corr-func-neq0}
\langle V_{j}(z_1),V_{j}(z_2)\rangle\neq0
\end{equation}
for any $z_1$ and $z_2$, then the computed $\lambda_j$ is independent of $z_0$. For pure edge modes, Eq.~(\ref{eq:corr-func-neq0}) is satisfied because
\begin{align*}
&|\langle V_{j}(Z_1),V_{j}(Z_2)\rangle|=|\langle b^S(Z_1),b^S(Z_2)\rangle|\\
=&\frac{(1-\rho^2|\gamma(Z_1)|^2)^{1/2}(1-\rho^2|\gamma(Z_2)|^2)^{1/2}}{|1-\rho^2\gamma(Z_1)\gamma^\ast(Z_2)|}\neq0
\end{align*}
for any $Z_1$ and $Z_2$, where Eq.~(\ref{statsoln}) and $|\rho\gamma(Z)|<1$ are used. If Eq.~(\ref{eq:corr-func-neq0}) is not satisfied, then $\lambda_j$ may be regarded as intrinsically multi-valued. Since in either case $z_0$ may be chosen arbitrarily, in the following we simply choose  $f=V_{j}(0)$.

\begin{figure}
\centering
\begin{tabular}{ccc}
\includegraphics[width=.32\textwidth]{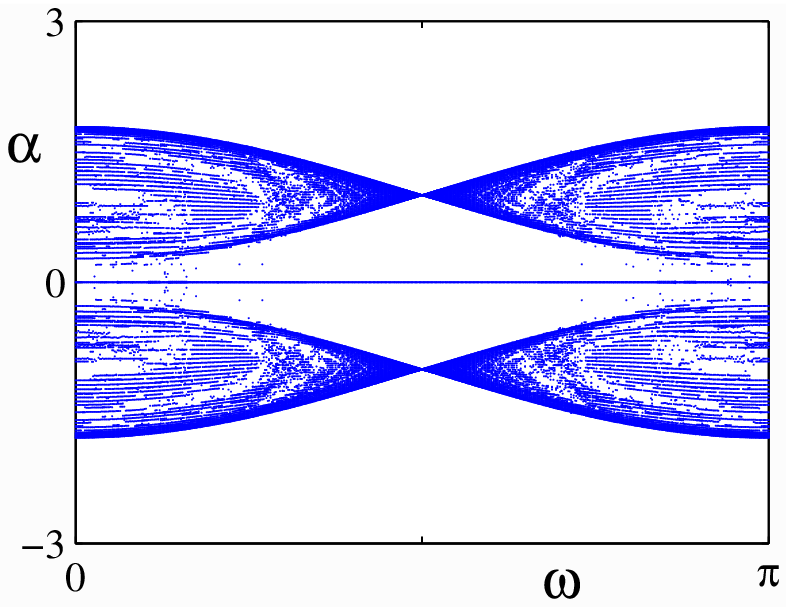} &
\includegraphics[width=.32\textwidth]{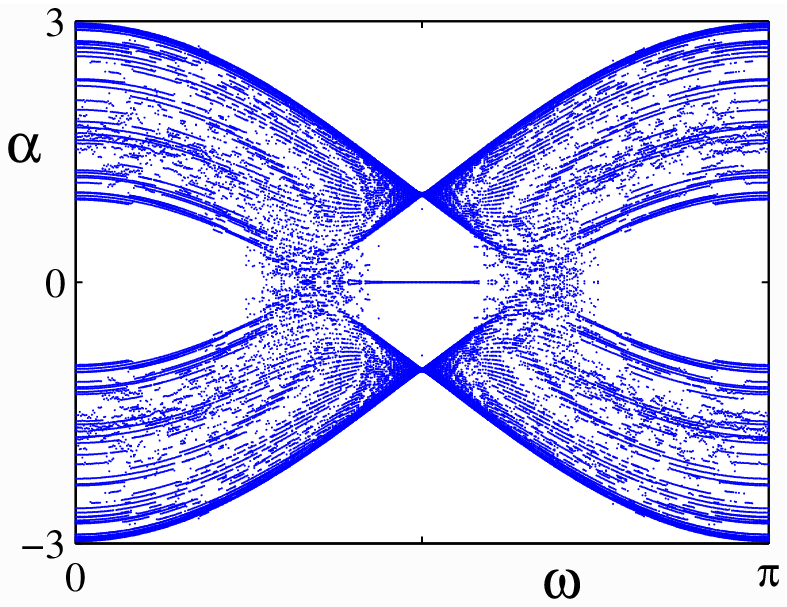} &
\includegraphics[width=.32\textwidth]{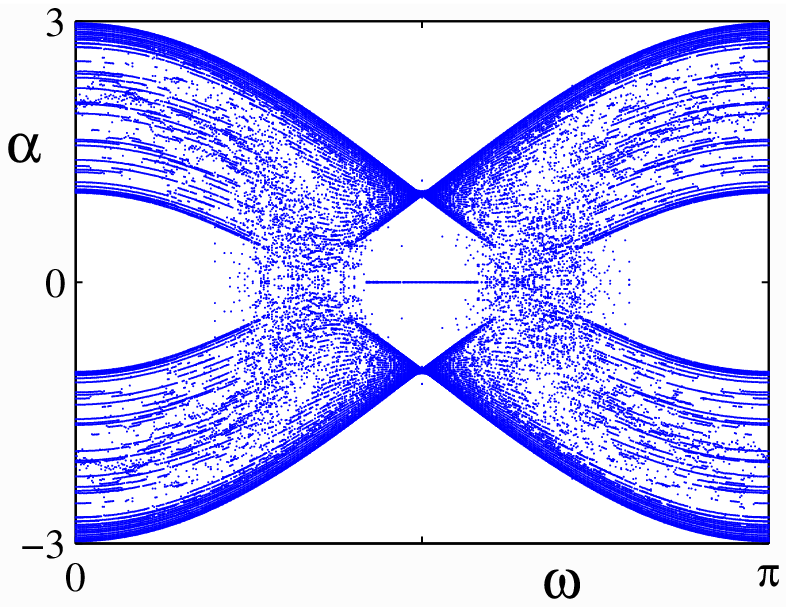} \\
(a) & (b) & (c)
\end{tabular}
\caption{The full Floquet spectra computed at the same parameters as in Fig.~\ref{fig:EDGE_disp_pure}.}
\label{fig:EDGE_slow_disp_full}
\end{figure}

Figure~\ref{fig:EDGE_slow_disp_full} shows the full Floquet spectra computed at the same parameters as in Fig.~\ref{fig:EDGE_disp_pure}. Note that Fig.~\ref{fig:EDGE_disp_pure} is a blowup of Fig.~\ref{fig:EDGE_slow_disp_full} around $\alpha=0$, such that the edge modes plotted in Fig.~\ref{fig:EDGE_disp_pure}  appear essentially flat in Fig.~\ref{fig:EDGE_slow_disp_full}. For either $\rho<1/2$ or $\rho>1/2$, the overall structure of the spectrum is similar to the case where the pseudo-field ${\bf A}$ is absent~\cite{abl5}, though in our case the bulk spectrum no longer consists of regular bands. Despite this loss of regularity, it is interesting to note that the bulk spectrum is non-ergodic and tends to avoid certain regions on the $(\omega,\alpha)$-plane.

\subsection{Length Scales Associated with Quasi-Edge States}\label{sec:length-scale}
The scales of the problem suggest that both pure and quasi- edge modes might be observable in experiments. A discussion of potential scales for  such an observation follows. In the introduction it was shown that the typical length scale in the longitudinal direction is $z_*=2k_0l^2$, where $k_0= \frac{2 \pi n_0}{\lambda}$, $\lambda$ being the input wavelength.  Typical values (cf.~\cite{rechts3}) are $\lambda=633$ nm, $n_0=1.5$, $l=15$ $\mu$m; this leads to $z_*= 6.75\times10 ^{-3}$ m.

For the pseudo-field Eq. (\ref{eq:A-rechts})
\begin{align*}
{\bf A}&=(A_1(z),A_2(z))= \kappa (\sin{\Omega z}, -\cos{\Omega z}) \\
&= \kappa (\sin{Z}, -\cos{Z}),\quad Z=\epsilon z,
\end{align*}
taking $\Omega z_*= \epsilon$ where $\epsilon=0.1$ leads  to $\Omega= 14.8$ rad/m and the period $T= \frac{2 \pi}{\Omega}=42$ cm. Thus using the parameters in Fig.~\ref{fig:EDGE_slow_exic}(b), namely a perfect HC lattice with $\rho=1$, the strength of the pseudo-field $\kappa=0.3$ and an input transverse wavenumber $\omega = \frac{2 \pi}{3}$, the edge mode disintegrates at $z_d=T/4\approx 10.5$ cm.

It is useful to compare this length scale with the fast evolution problem discussed in \cite{rechts3,ACM14fast}. The parameters used in this case were  $\epsilon =0.24$, and $\Omega z_*= \frac{1}{\epsilon}$, which  leads to  $\Omega= 628$ rad/m and the period $T= \frac{2 \pi}{\Omega}=1$ cm.  As in the slow evolution case, a perfect HC lattice with  $\rho=1$ was taken, while the strength of the pseudo-field was $\kappa=1.4$.
\section{Nonlinear Two-Dimensional Localized Edge Modes}
Importantly, nonlinear edge modes can also be constructed via the same asymptotic analysis used in Sec.~\ref{4}. In this case, in the presence of weak nonlinearity where $\sigma = \epsilon \tilde{\sigma}$, the Fredholm condition (\ref{Fredholm}) associated with the edge solution (\ref{envelope_eq}--\ref{statsoln}) leads to the following equation for the envelope $C=C(Z;\omega)$
\begin{equation}\label{eq:nlin-fC}
i\p_{Z}\fC = \alpha_l(Z;\omega) \fC - \tilde{\sigma}  \alpha_{nl}(Z;\omega) |\fC|^{2}\fC,
\end{equation}
where $\alpha_{nl}(Z;\omega) = \|b^{S}(Z)\|_{4}^{4}/\|b^S(Z)\|_2^2$ with
\[
\left|\left|b^S\right|\right|_2^2= \sum_{n=0}^{\infty}\left|b_{n}^{S} \right|^{2},  \quad    \left|\left|b^{S} \right|\right|_{4}^{4} = \sum_{n=0}^{\infty}\left| b_{n}^{S} \right|^{4}.
\]
We can reconstruct the approximation to $b_{mn}$ via
\begin{equation}\label{eq:bmn}
b_{mn}=C(Z, \omega)e^{i \omega m} b_n^S(Z).
\end{equation}

Fixing the time $Z$, we define the frequency interval $\mathcal{I}_{\omega}(Z)$ to be
\[
\mathcal{I}_{\omega}(Z) = \left\{\omega : \mbox{Eq.}~\eqref{eq:omega-varphi-theta}~\mbox{is satisfied} \right\}.
\]
In the narrow band approximation with $\omega$ near any given $\omega_0\in\mathcal{I}_\omega(Z)$, the solution $C$ represents an envelope function with carrier wavenumber $\omega_0$. To describe its dynamics, we first expand $\alpha_l(Z;\omega)$ and $\alpha_{nl}(Z;\omega)$ around $\omega_0$.  We then replace $\omega-\omega_0$ by $-i \nu \partial_y$, where $\nu$ is the width around $\omega_0$, or the inverse width of the envelope in physical space; see also \cite{ACM14fast}. With this Eq.~(\ref{eq:nlin-fC}) transforms to the following equation for the envelope $C$
\begin{align}
i\p_Z C =& \left[\sum_{j=0}^2\frac{\alpha_l^{(j)}(Z;\omega_0)}{j!}(-i \nu \partial_y)^j+O(\nu^3)\right] C\nonumber\\
& -\tilde{\sigma} \left[\alpha_{nl}(Z;\omega_0)+O(\nu)\right] |C|^2C,\label{eq:nlin-pC}
\end{align}
where $\alpha_l^{(j)}(Z;\omega_0)$ denotes the $j$-th derivative of $\alpha_l(Z;\omega)$ with respect to $\omega$ at $\omega=\omega_0$. At leading order, Eq.~(\ref{eq:nlin-pC}) reduces to the following nonconstant coefficient nonlinear Schr\"{o}dinger (NLS) equation
\begin{equation}\label{eq:NLS}
i\p_{\tilde{Z}} \tilde{C}   +  \frac{\alpha_l''(Z;\omega_0)}{2}   \tilde{C}_{YY} + \sigma_{eff}(Z;\omega_0)  |\tilde{C}|^{2} \tilde{C}=0,
\end{equation}
where
\begin{align*}
&C= \tilde{C}(\tilde{Z},Y) \exp{(-i \int_0^Z\alpha_l(t;\omega_0)dt)},\\
&Y= y- \nu\int_0^Z\alpha_l'(t;\omega_0)dt,\\
&\tilde{Z}= \nu^2 Z, \quad \sigma_{eff}(Z;\omega_0)=\tilde{\sigma} \alpha_{nl}(Z;\omega_0)/\nu^2.
\end{align*}
Equation~(\ref{eq:NLS}) is maximally balanced when $\sigma_{eff}(Z;\omega_0)=O(1)$. At time $Z$, Eq.~(\ref{eq:NLS}) is focusing (defocusing) if $\alpha_l''(Z;\omega_0)\sigma_{eff}(Z;\omega_0)>0~ (<0$). Since the periodic average of $\alpha_l(Z;\omega)$ in $Z$ is $ \bar{\alpha} (\omega)$, and $\sigma_{eff}(Z;\omega)$ always has the same sign as $\tilde{\sigma}$, Eq.~(\ref{eq:NLS}) is on average focusing (defocusing)  if $ \bar{\alpha} ''(\omega_0)\tilde{\sigma}>0 ~(<0)$. In the on average focusing case, the NLS equation is expected to contain solitons, and so the 2D discrete system Eqs.~(\ref{eq:abmn-a}--\ref{eq:abmn-b}) is expected to contain edge solitons. In the on average defocusing case, dispersion dominates on average, so no soliton is expected.

\begin{figure}
\centering
\begin{tabular}{cc}
\includegraphics[width=0.32\textwidth]{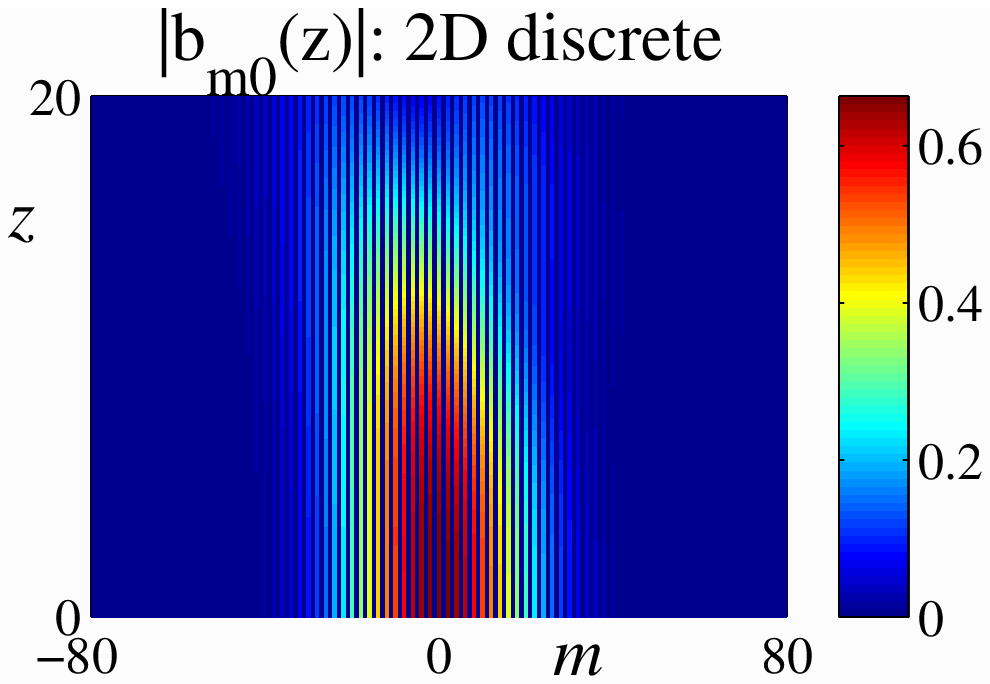} &
\includegraphics[width=0.32\textwidth]{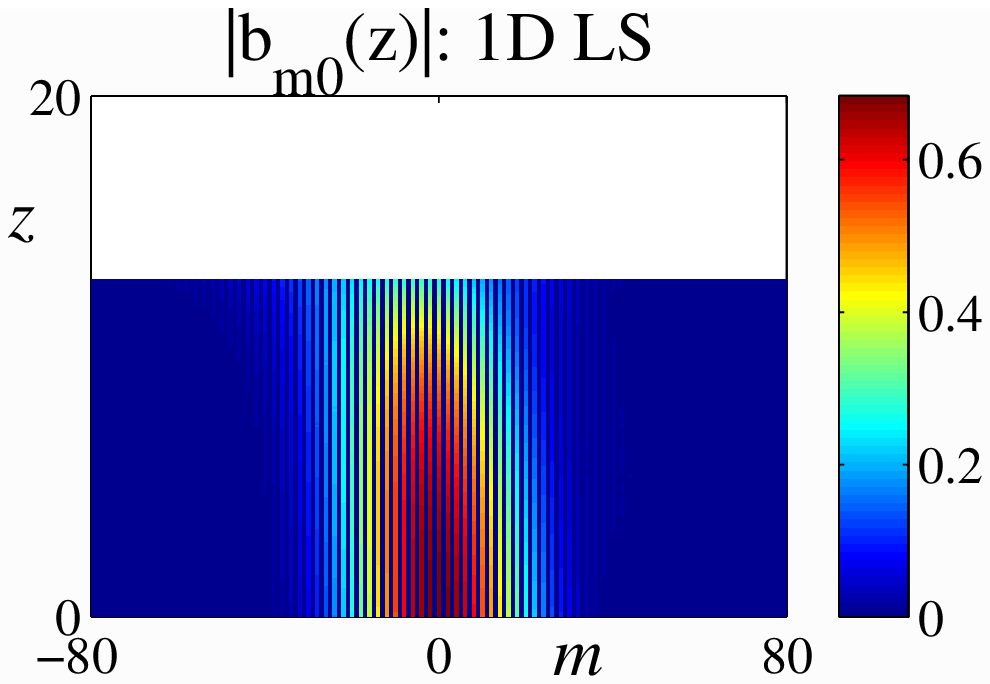}\\
(a) & (b)
\end{tabular}
\caption{Plot of the time evolution of the 2D discrete system Eqs.~(\ref{eq:abmn-a}--\ref{eq:abmn-b}) with $\sigma=0$ (panels (a)) and the 1D LS equation (\ref{eq:nlin-pC}) with $\tilde{\sigma}=0$ (panel (b)) in terms of $|b_{m0}(z)|$ (at the edge). The parameters are $(\rho,\kappa)=(1,0.3)$, $\epsilon=2\pi/60$, and $\omega_0=2\pi/3$ as in Fig.~\ref{fig:EDGE_slow_exic}(b), and the envelope width is $\nu=0.1$. Periodic boundary conditions in $m$ are used. Note in panel (b) the computation of the 1D LS equation stops at $z\approx z_+=15$ where the edge state delocalizes.} \label{fig:EDGE_slow_mixed_2d}
\end{figure}

To test these predictions, we solve the 2D discrete system Eqs.~(\ref{eq:abmn-a}--\ref{eq:abmn-b}) numerically using the initial condition
\begin{equation}
a_{mn}(Z=0)=0,\quad b_{mn}(Z=0) = \int_{\mathcal{I}_\omega(Z=0)} \hat{b}(\omega) b_{n}^{S}(Z=0;\omega)e^{im\omega} d\omega,
\label{envelope}
\end{equation}
with a narrow envelope
\[
\hat{b}(\omega) = \frac{e^{-(\omega-\omega_0)^2/\nu^2}}{ \int_{\mathcal{I}_\omega(Z=0)} e^{-(\omega-\omega_0)^2/\nu^{2}} d\omega },
\]
and compare the results with $b_{mn}$ reconstructed from numerical solutions of the 1D NLS equation~(\ref{eq:nlin-pC}) with the initial condition
\begin{equation}
C(Z=0,y)=\int_{\mathcal{I}_\omega(Z=0)} \hat{b}(\omega)e^{iy(\omega-\omega_0)/\nu} d\omega,
\end{equation}
where we note that $C(Z=0,y=0)=1$. Throughout this section we take ${\bf A}(Z)$ to be the circular pseudo-field Eq.~(\ref{eq:A-rechts}).
In Fig.~\ref{fig:EDGE_slow_mixed_2d}, we compare linear ($\sigma=0$) quasi-edge modes found from the full 2D discrete system (Fig.~\ref{fig:EDGE_slow_mixed_2d} (a)) to those found from the 1D linear ($\tilde{\sigma}=0$) Schr\"{o}dinger (LS) equation (Fig.~\ref{fig:EDGE_slow_mixed_2d} (b)).  The comparison of results is shown in terms of $|b_{m0}(z)|$. For the 1-D LS equation, the following modification of Eq.~(\ref{eq:bmn}),
\begin{equation}\label{eq:bmn-y}
b_{mn}=C(Z,y)e^{i\omega_0y/\nu}b_n^S(Z,\omega_0),
\end{equation}
is used to reconstruct $b_{mn}$ with $C$ satisfying the LS equation. The parameters are chosen to agree with Fig.~\ref{fig:EDGE_slow_exic}(b), such that the 1D stationary mode $b_n^S(Z)$ disintegrates into the bulk at $z=z_+\approx15$. As shown in Fig.~\ref{fig:EDGE_slow_mixed_2d}(a), the 2D localized mode with a narrow envelope $\nu=0.1$ also disintegrates into the bulk around $z=z_+$. As shown in Fig.~\ref{fig:EDGE_slow_mixed_2d}(b), this 2D evolution before $z=z_+$ is well described by the 1D LS equation. After $z=z_+$, the 1D LS equation is no longer valid because $\alpha_l''(Z;\omega_0)$ blows up at $z=z_+$. For $z>z_+$, the 2D evolution reveals that most power is concentrated in the bulk and distributed between the $a$ and $b$ lattice sites. Interestingly, the 2D mode, though small, still remains localized in $m$ for some time even after scattering into the bulk begins at $z=z_+$.

\begin{figure}
\centering
\begin{tabular}{cc}
\includegraphics[width=0.32\textwidth]{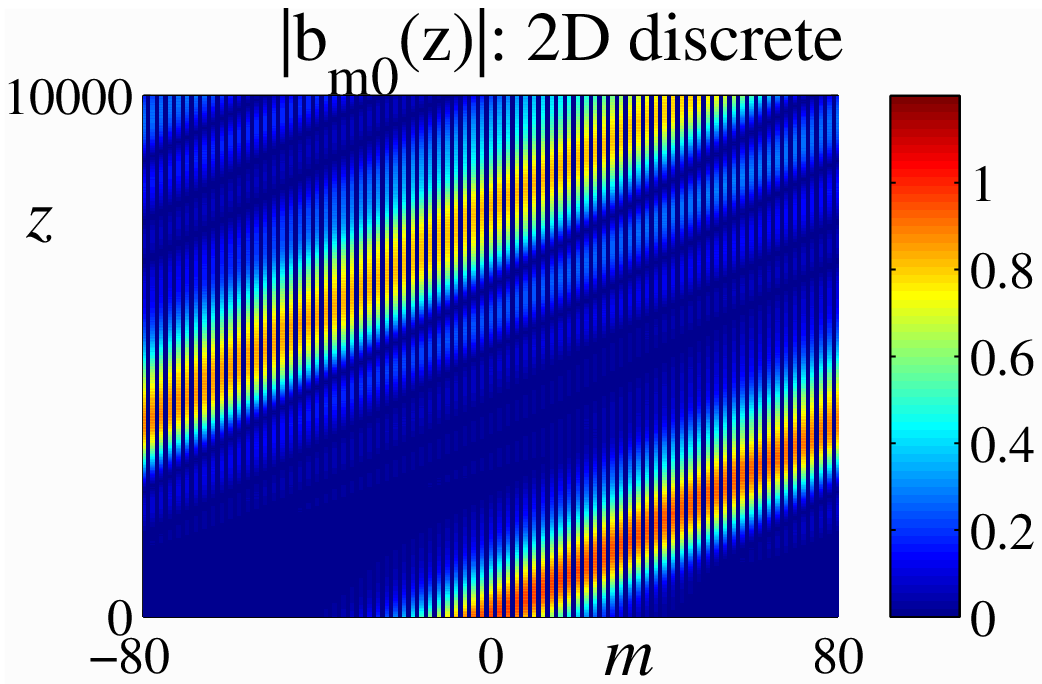} &
\includegraphics[width=0.32\textwidth]{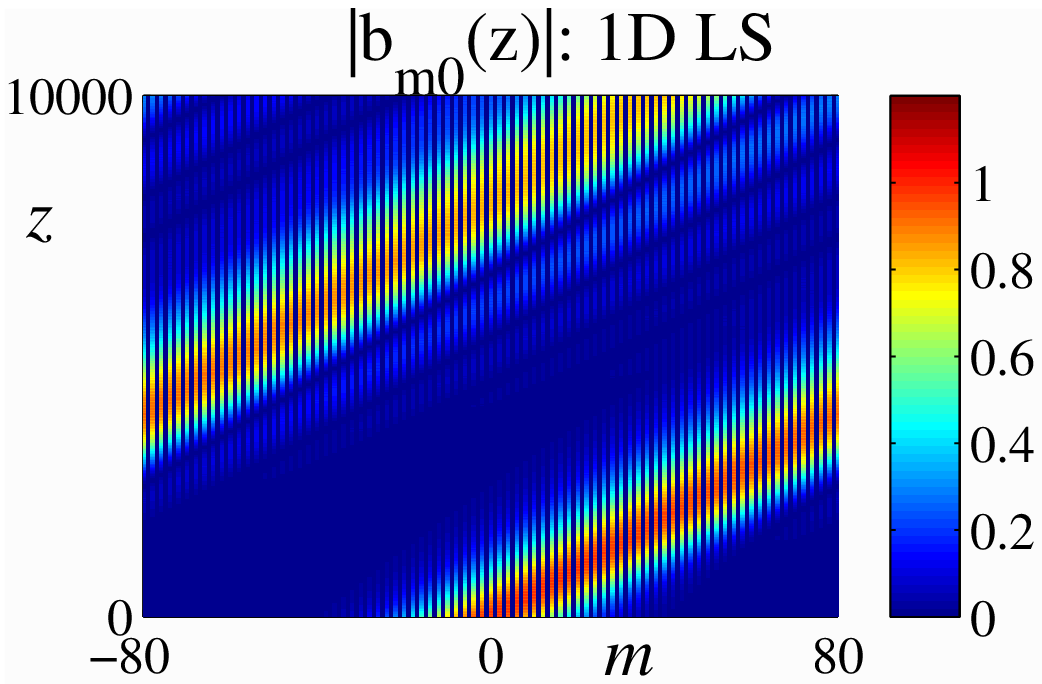}\\
(a) & (b) \\
\includegraphics[width=0.32\textwidth]{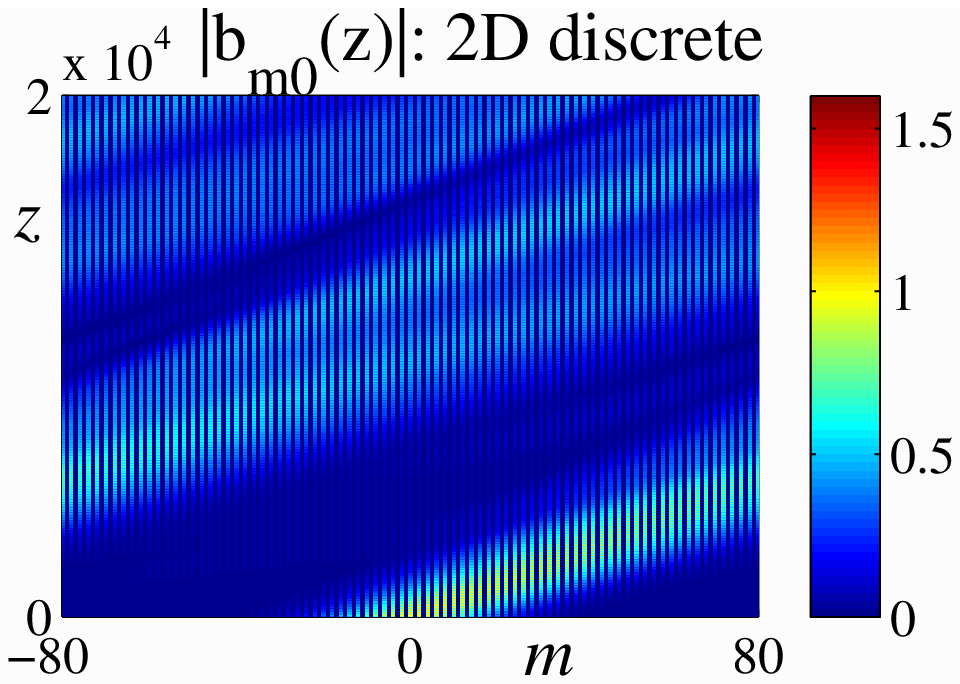} &
\includegraphics[width=0.32\textwidth]{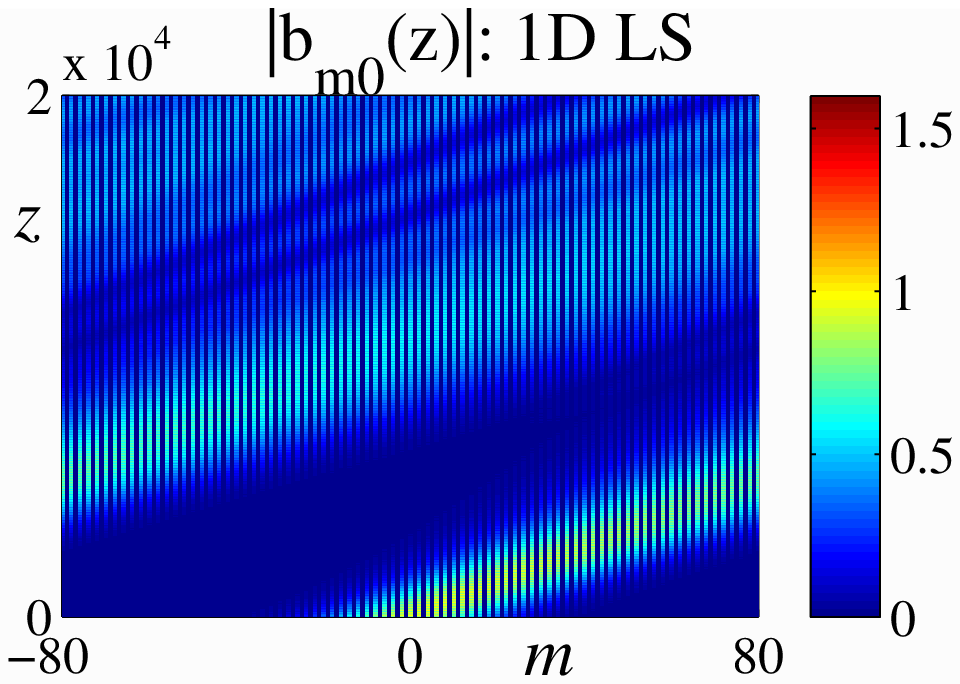}\\
(c) & (d) \\
\includegraphics[width=0.32\textwidth]{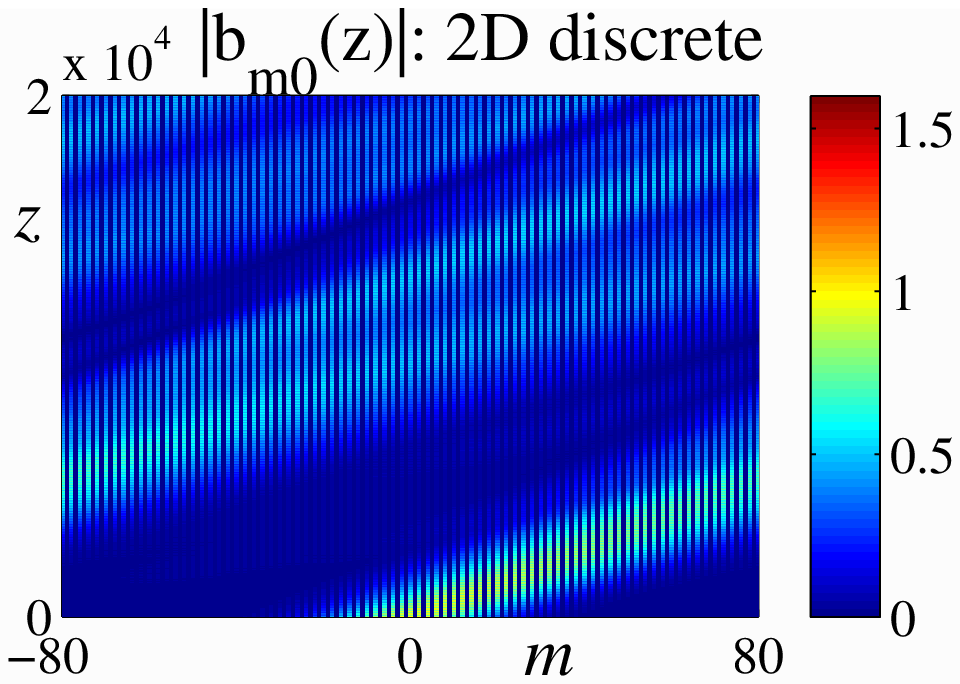} &
\includegraphics[width=0.32\textwidth]{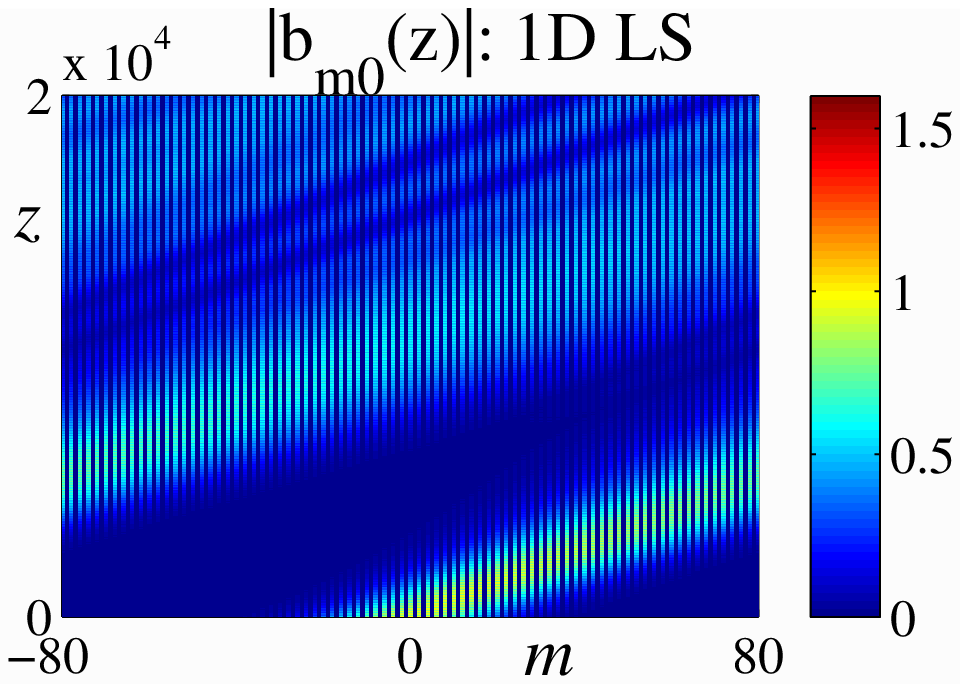}\\
(e) & (f)
\end{tabular}
\caption{Plot of the time evolution of the 2D discrete system Eqs.~(\ref{eq:abmn-a}--\ref{eq:abmn-b}) with $\sigma=0$ (panels (a,c,e)) and the 1D LS equation (\ref{eq:nlin-pC}) with $\tilde{\sigma}=0$ (panels (b,d,f)), shown in terms of $|b_{m0}(z)|$ (at the edge). The parameters for panels (a--b) agree with Fig.~\ref{fig:EDGE_slow_exic}(a) with $\nu=0.1$. The parameters for panels (c--f) agree with Fig.~\ref{fig:EDGE_slow_disp_rho0d51}(a) with $(\omega_{0},\nu)$: (c,d) $(3\pi/8,0.1)$; (e,f) $(5\pi/8,0.1)$. The edge state persists over a long distance.  Periodic boundary conditions in $m$ are used.} \label{fig:EDGE_slow_line}
\end{figure}

\begin{figure}
\centering
\begin{tabular}{cc}
\includegraphics[width=.32\textwidth]{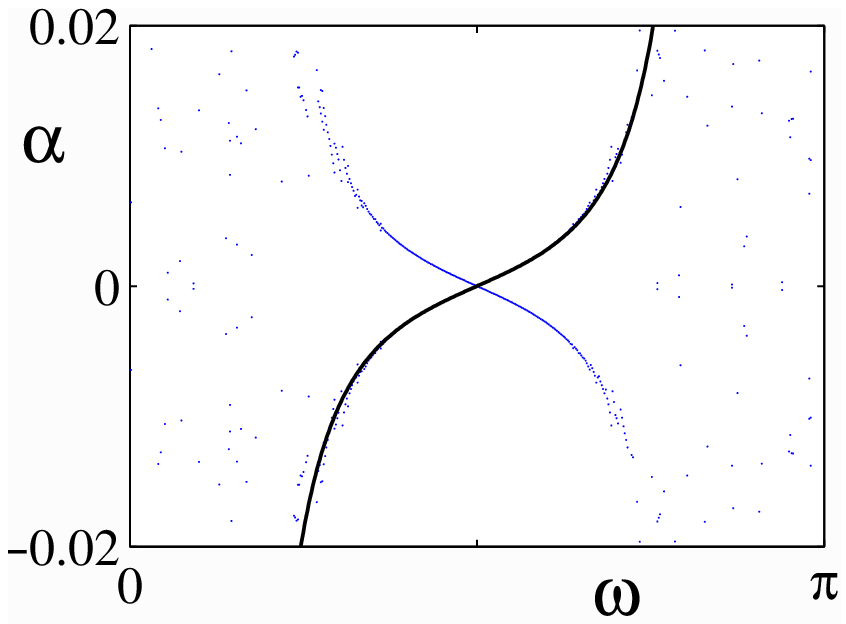} &
\includegraphics[width=.32\textwidth]{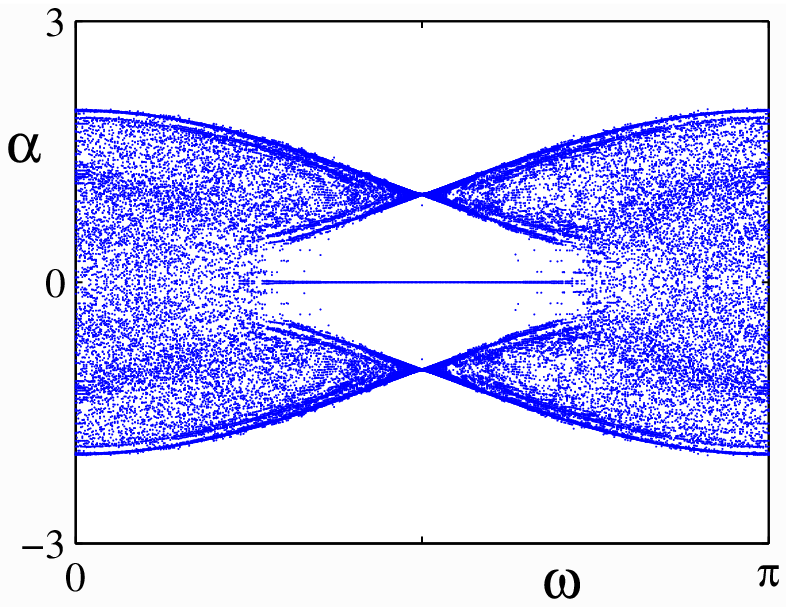} \\
(a) & (b)
\end{tabular}
\caption{The dispersion relation of pure edge modes (panel (a)) and the full Floquet spectrum (panel (b)) computed at $\kappa=0.5$ and $\rho=0.51$, using $40$ lattice sites and $\epsilon=2\pi/100$. The blue curve represents pure edge modes computed numerically from Eqs.~(\ref{eq:abn-ori-a}--\ref{eq:abn-ori-b}). The black curve shows the asymptotic prediction Eq.~(\ref{eq:talpha-omega}).}
\label{fig:EDGE_slow_disp_rho0d51}
\end{figure}

In Fig.~\ref{fig:EDGE_slow_line}, we compare linear ($\sigma=0$) pure edge modes found from the full 2D discrete system to those found from the 1D LS equation. As before, the comparison of results is shown in terms of $|b_{m0}(z)|$, the panels (a,c,e) show the solutions of the 2D discrete system and the panels (b,d,f) show the solutions of the 1D LS equation with Eq.~(\ref{eq:bmn-y}) used to reconstruct $|b_{m0}(z)|$. The parameters for panels (a--b) are chosen to agree with Fig.~\ref{fig:EDGE_slow_exic}(a). The parameters for panels (c--f) are chosen to agree with Fig.~\ref{fig:EDGE_slow_disp_rho0d51}(a), which has a wider localization interval $\mathcal{I}_p$ than Fig.~\ref{fig:EDGE_disp_pure}(b); see also the corresponding full Floquet spectrum in Fig.~\ref{fig:EDGE_slow_disp_rho0d51}(b). The value of $\omega_0$ is chosen such that $\bar{\alpha} ''(\omega_0)=0$ for panels (a,b), $\bar{\alpha} ''(\omega_0)<0$ for panels (c,d), and $\bar{\alpha} ''(\omega_0)>0$ for panels (e,f). In all cases, it can be seen that the localized mode is eventually destroyed by dispersion after sufficient evolution. In the $\bar{\alpha} ''(\omega_0)=0$ case, the third derivative term in Eq.~(\ref{eq:nlin-pC}) should be kept, which leads to a 1D zero-dispersion LS equation. In this case the mode disperses more gradually. As expected, the 1D LS equation reproduces the time evolution of the 2D discrete system well up to $z\sim1/(\epsilon\nu^3)$ for panel (a) and $z\sim1/(\epsilon\nu^2)$ for panels (c,e). Beyond this time scale, the 2D evolution becomes somewhat weaker than predicted by the LS equation due to the transfer of power from $b$ to $a$ as well as higher-order dispersion effects. Nevertheless the edge state persists over a long distance.

\begin{figure}
\centering
\begin{tabular}{cc}
\includegraphics[width=0.32\textwidth]{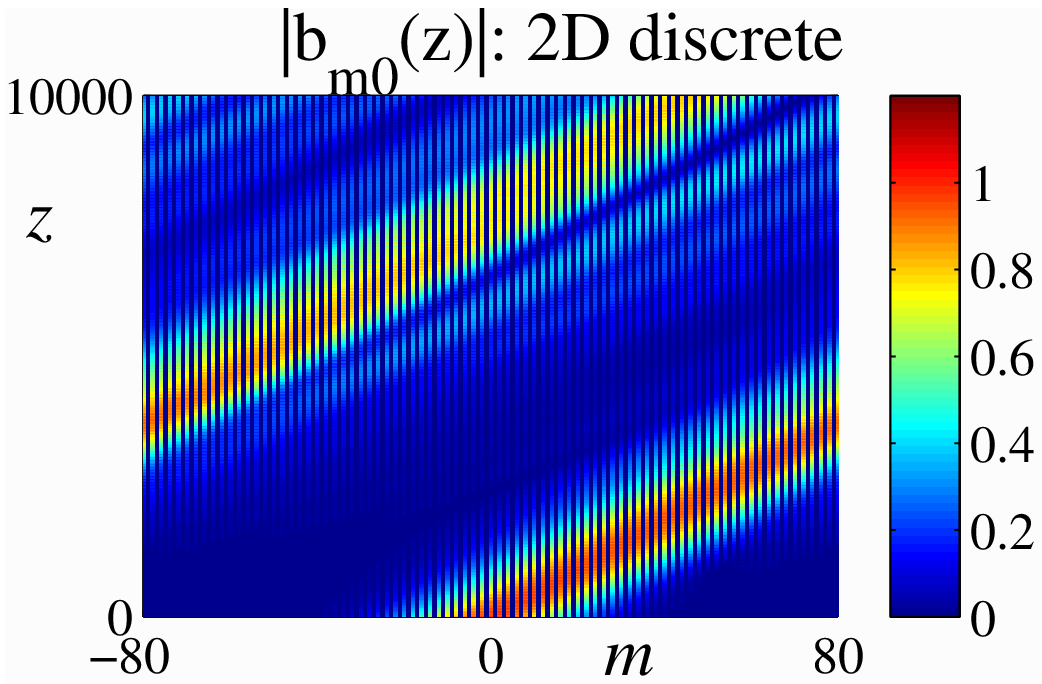} &
\includegraphics[width=0.32\textwidth]{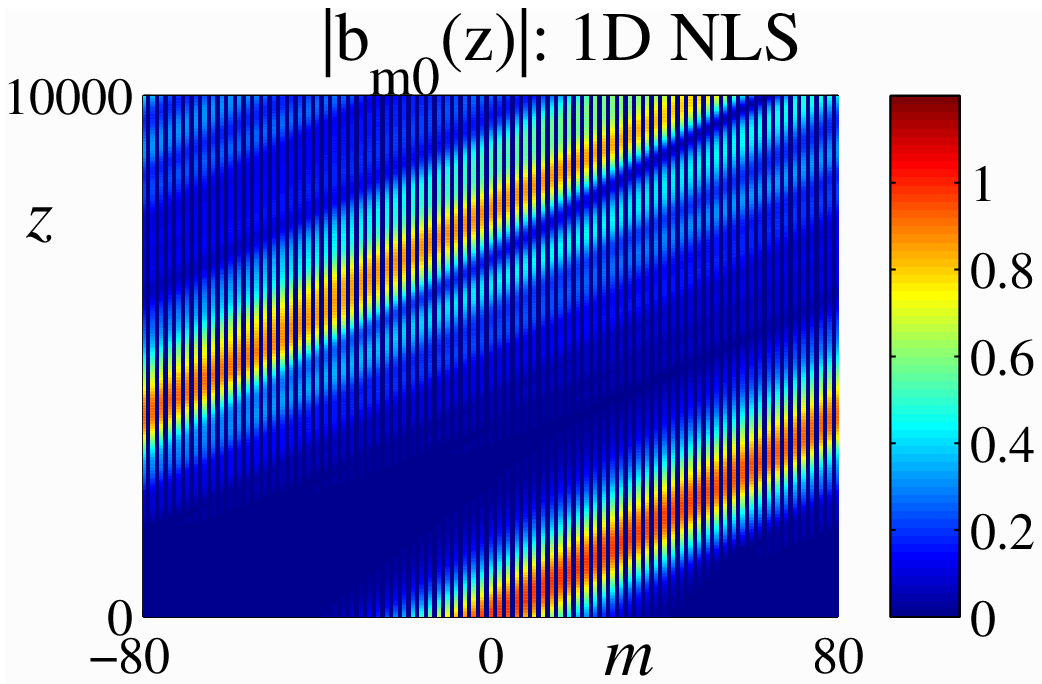}\\
(a) & (b) \\
\includegraphics[width=0.32\textwidth]{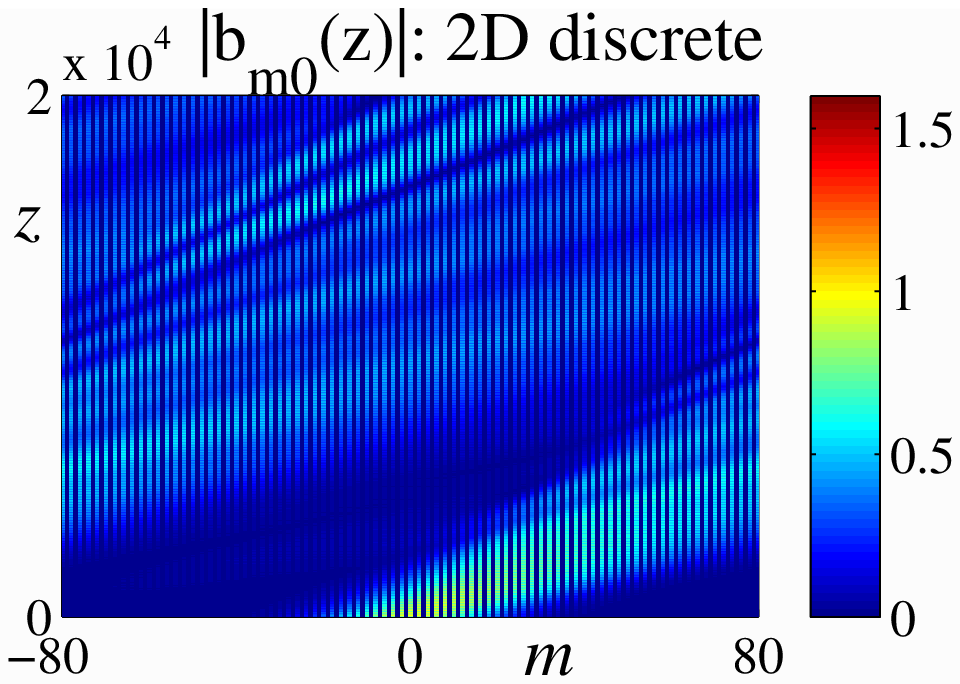} &
\includegraphics[width=0.32\textwidth]{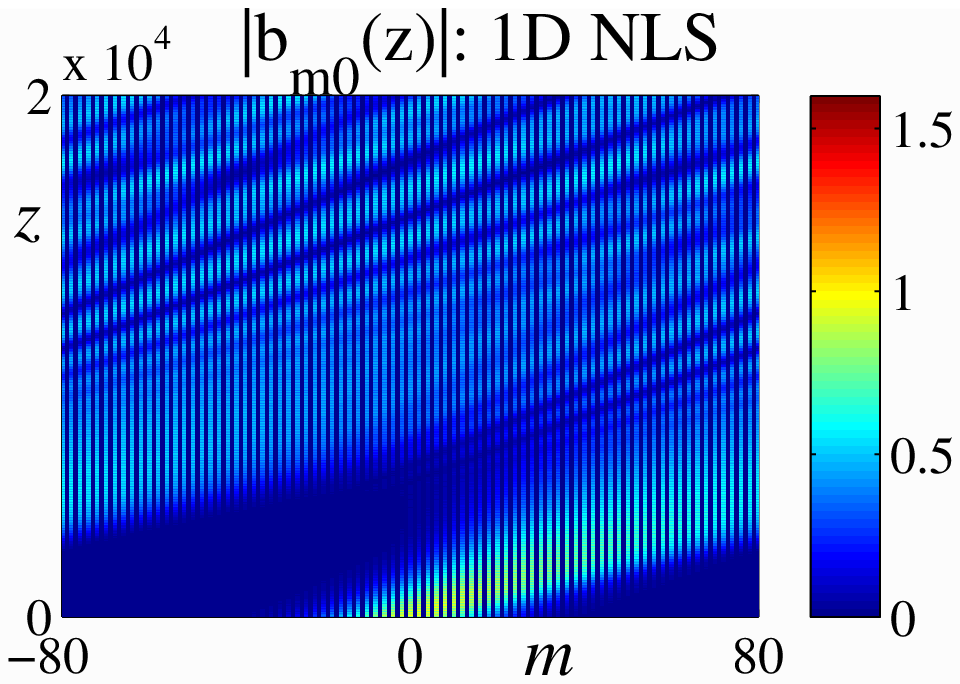}\\
(c) & (d) \\
\includegraphics[width=0.32\textwidth]{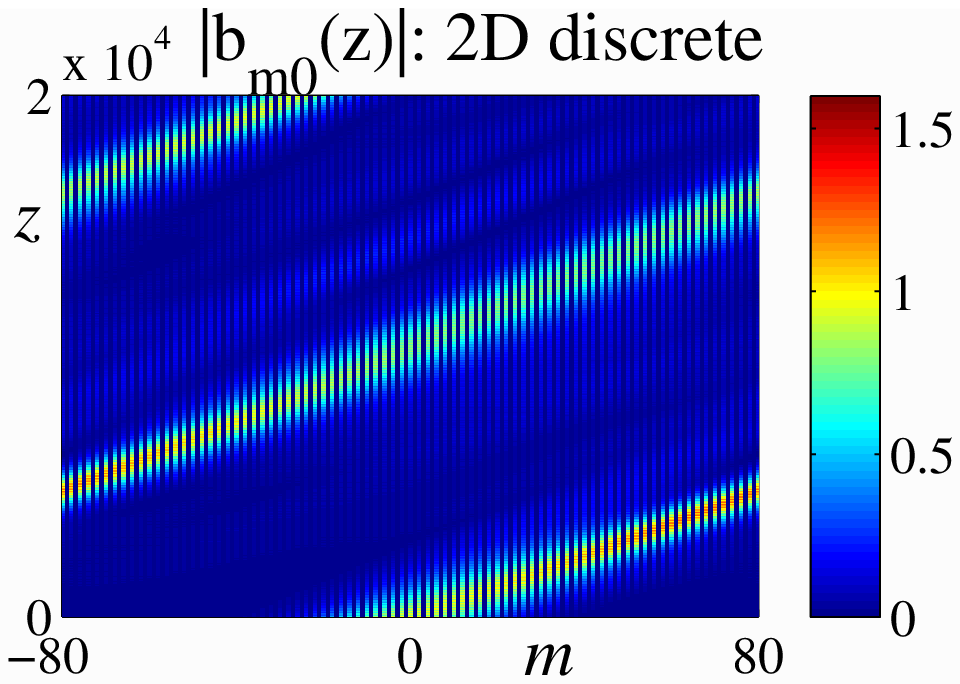} &
\includegraphics[width=0.32\textwidth]{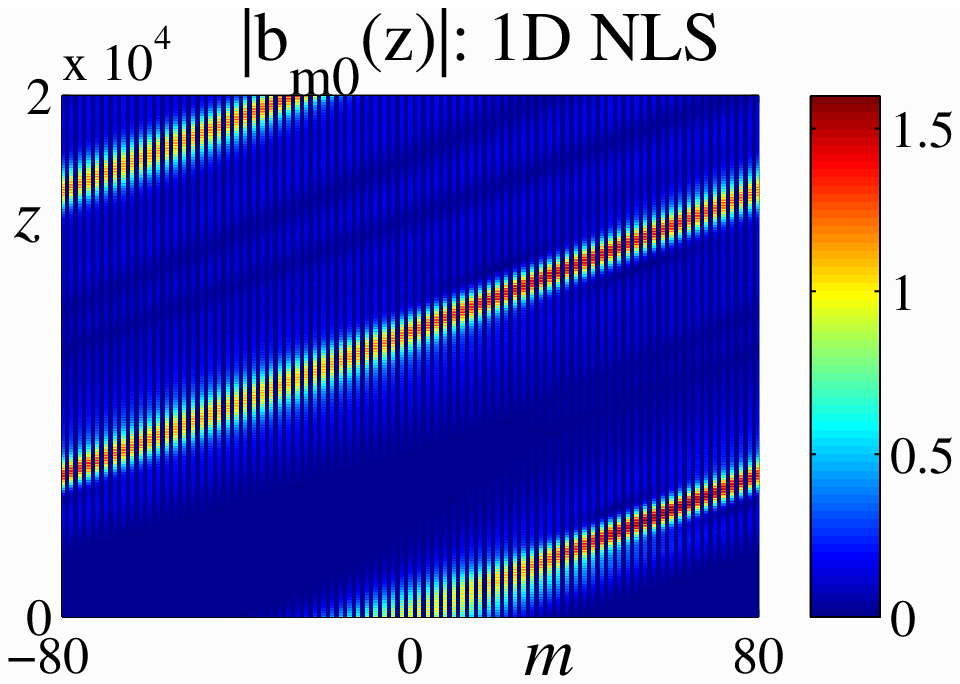}\\
(e) & (f)
\end{tabular}
\caption{Plot of $|b_{m0}(z)|$ (at the edge) at the same parameters as in Fig.~\ref{fig:EDGE_slow_line} but with $\sigma$: (a,b) $5\times10^{-4}$; (c--f) $2\times10^{-3}$.
Remarkably, in panel (e) the edge soliton persists until at least  $z =2\times10^4$ and has the signature of a nonlinear topological edge state.}
\label{fig:EDGE_slow_nlin}
\end{figure}

Figure~\ref{fig:EDGE_slow_nlin} shows the nonlinear evolution at the same parameters as Fig.~\ref{fig:EDGE_slow_line} but with $\sigma\neq0$. As shown in Fig.~\ref{fig:EDGE_slow_nlin}(a,b), when the NLS equation has third order dispersion due to $ \bar{\alpha} ''(\omega_0)=0$, weak nonlinearity enhances dispersion somewhat. As shown in Fig.~\ref{fig:EDGE_slow_nlin}(c,d), when the NLS equation is primarily defocusing due to $ \bar{\alpha} ''(\omega_0)<0$, weak nonlinearity also enhances dispersion. As shown in Fig.~\ref{fig:EDGE_slow_nlin}(e,f), when the NLS equation is primarily focusing due to $ \bar{\alpha} ''(\omega_0)>0$, weak nonlinearity enhances localization. In the last case, an edge soliton is formed which remains localized  over very long distances; we see from the figure that the edge soliton remains intact at least until $z=2\times10^4$.  Due to the $z$-dependence of the coefficients of the NLS equation, the edge soliton exhibits slow modulation in its amplitude and width.

We remark that the $z$-dependent NLS equation exhibits various other interesting dynamics in suitable parameter regimes, such as the splitting of a single soliton into two solitons which propagate at different speeds. As in the linear case, the 2D evolution becomes somewhat weaker than predicted by the NLS equation beyond the time scale $z\sim1/(\epsilon\nu^3)$ for panels (a,b) and $z\sim1/(\epsilon\nu^2)$ for panels (c--f). This effect is especially apparent in the amplitude of the edge solitons shown in Fig.~\ref{fig:EDGE_slow_nlin}(e,f). Despite this slow loss of amplitude, it is remarkable that the edge soliton propagates at a constant speed for such a long distance. This absence of backscattering in the presence of nonlinearity suggests that the edge soliton is indeed topologically protected in the same parameter regime as  topologically protected linear modes. But in fact, they remain localized for a much longer distance than the linear case.  This shows that nonlinearity enhances the robustness of edge modes.

\section{Conclusion}
In this paper, a method is developed which describes the propagation of edge modes in a semi-infinite honeycomb lattice in the presence of a periodically and relatively slowly varying pseudo-field with weak nonlinearity. Two types of edge modes are found, referred to respectively as pure and quasi-edge modes. Pure edge modes remain localized for the entire period, while quasi-edge modes remain localized for only part of the period. In the linear case, the dispersion relations of pure edge modes indicate that some modes may exhibit topological protection. With weak nonlinearity included, it is shown that in the narrow band approximation, a time-dependent NLS equation is obtained. This NLS equation admits solitons, and they are found to be part of the long time nonlinear evolution under suitable circumstances. These 1D NLS solitons correspond to true edge solitons propagating on the edge of the semi-infinite honeycomb lattice. Finally, over very long distances, with certain choices of parameters consistent with the notion of topological protection as indicated by the linear dispersion relation, localized nonlinear edge modes in the focusing case  are found to also be  immune from backscattering. On the other hand when the NLS equation is defocusing significant dispersion occurs.
\section*{Acknowledgements}
This research was partially supported by the U.S. Air Force Office of Scientific Research, under grant FA9550-12-1-0207 and by the NSF under grants DMS-1310200, CHE 1125935.

\appendix
\section{Perturbation Method--Fredholm Condition}\label{pert}
In Section \ref{discrete eq}, the basic multiple scales perturbation procedure was formulated, and the leading order solution was determined. In this Appendix, we carry out the procedure from that point.  At $O(\epsilon)$ the perturbation equation is
\begin{align}
i\p_{z}a^{(1)}_{n}  + e^{i\bdd\cdot{\bf A}}\mathcal{L^-}b^{(1)}_n= F^{(0)}_{n,1}, \label{eq:abn2-ori-a} \\
i\p_{z}b^{(1)}_{n}  +  e^{-i\bdd\cdot{\bf A}}\mathcal{L^+}a^{(1)}_n= F^{(0)}_{n,2} , \label{eq:abn2-ori-b}
\end{align}
where
\begin{align}
 F^{(0)}_{n,1} =-\left( ia^{(0)}_{n,Z}+\tilde{\sigma} |a^{(0)}_{n}|^{2}a^{(0)}_{n}\right), \label{eq:abn3-ori-a} \\
 F^{(0)}_{n,2}=  - \left( ib^{(0)}_{n,Z}+\tilde{\sigma} |b^{(0)}_{n}|^{2}b^{(0)}_{n}\right). \label{eq:abn3-ori-b}
\end{align}

In order for functions $a^{(1)}_n,b^{(1)}_n$ to have decaying solutions at infinity, the following Fredholm condition must  be satisfied
\begin{equation}
\sum_{n=0}^{\infty} F^{(0)}_{n,2} (b_{n}^S(Z))^*=0.
\label{Fredholm}
\end{equation}
The Fredholm condition (\ref{Fredholm}) is obtained from the identity
\begin{align*}
e^{-i\bdd\cdot{\bf A}}\mathcal{L^+}a^{(1)}_n (b^{(0)}_n)^*- (e^{i\bdd\cdot{\bf A}}\mathcal{L^-}b^{(0)}_n)^*a^{(1)}_n =\\
e^{-i\bdd\cdot{\bf A}} \rho \gamma \Delta_n(a^{(1)}_n (b^{(0)}_{n-1})^*)= F^{(0)}_{n,2} (b^{(0)}_n)^*
\end{align*}
where $\Delta_n(G_n)=G_{n+1}-G_n$, summing over the lattice points and using the boundary conditions on $b_n^S$.  We shall only use the condition (\ref{Fredholm}) to determine the evolution of the leading order function $C(Z, \omega)$. In principle, one can solve for the decaying functions $a^{(1)}_n,b^{(1)}_n$ in order to obtain more accurate approximation, but going to higher order in the perturbation scheme is outside the scope of this paper.

\section{Asymptotic Behavior of Dispersion Relationships}\label{app:alpha-asym}
In this Appendix we show the asymptotic behavior (\ref{eq:alpha-asym-m}--\ref{eq:alpha-asym-p}) of dispersion relations in Case (II). To obtain Eq.~(\ref{eq:alpha-asym-m}), we first expand $\alpha_l(Z)$ around $Z=Z_-$ and $\omega=\omega_-$ to yield at leading order
\begin{align*}
\alpha_l(Z)&=-\frac{\varphi_+'(Z_-)}{1-4\rho^2\cos^2\left(\varphi_-''(Z_-)\delta Z^2/2-\delta\omega-\tilde{\theta}\right)}\\
&=-\frac{\varphi_+'(Z_-)}{2\sqrt{4\rho^2-1}\left(-\varphi_-''(Z_-)\delta Z^2/2+\delta\omega\right)},
\end{align*}
where $\delta Z\equiv Z-Z_-$ and $\delta\omega\equiv\omega-\omega_-$. Integration in $Z$ then yields
\begin{align*}
\bar{\alpha}=\frac{1}{T}\int_0^T\alpha_l(Z)dZ=-\frac{\varphi_+'(Z_-)}{2\sqrt{4\rho^2-1}}\frac{1}{\sqrt{-2\varphi_-''(Z_-)\delta\omega}}.
\end{align*}
Thus Eq.~(\ref{eq:alpha-asym-m}) is obtained by noting that $\varphi_+'(Z)=(\sqrt{3}/2)A_1'(Z)$ and $\varphi_-''(Z_-)<0$. The derivation of Eq.~(\ref{eq:alpha-asym-p}) is similar and omitted for brevity.

\end{document}